\documentclass[conference]{IEEEtran}

\usepackage{enumitem}
\usepackage{amsmath,amssymb,amsfonts}
\usepackage{cite}
\usepackage{graphicx}
\usepackage{xcolor}
\usepackage{subcaption}
\usepackage{tabularx}
\usepackage{multicol}
\usepackage{float}
\usepackage[export]{adjustbox}
\usepackage[noend,linesnumbered,ruled]{algorithm2e}
\SetAlFnt{\small}
\usepackage{booktabs}
\usepackage{array}
\usepackage[T1]{fontenc}
\usepackage{microtype}

\usepackage[normalem]{ulem}

\definecolor{darkyellow}{RGB}{255,168,67}
\definecolor{darkgreen}{RGB}{85,130,54}
\definecolor{lightgray}{RGB}{192,192,192}
\definecolor{purple}{RGB}{148,55,255}
\definecolor{orange}{RGB}{255,147,0}

\newcommand{\green}[1]{\textcolor{darkgreen}{#1}}

\newcommand{\flix}{\textbf{FliX}}
\newcommand{\flixst}{\textbf{FliX-ST}}

\newcommand{\btree}{\textbf{B-tree}}
\newcommand{\lsmtree}{\textbf{LSM-tree}}
\newcommand{\lsmu}{\textbf{LSMu}}

\newcommand{\slabhash}{\textbf{HT-Slab}}
\newcommand{\warphash}{\textbf{HT-Warpcore}}

\newcolumntype{L}[1]{>{\raggedright\let\newline\\\arraybackslash\hspace{0pt}}m{#1}}
\newcolumntype{C}[1]{>{\centering\let\newline\\\arraybackslash\hspace{0pt}}m{#1}}
\newcolumntype{R}[1]{>{\raggedleft\let\newline\\\arraybackslash\hspace{0pt}}m{#1}}

\providecommand{\Description}[1]{}
\usepackage{hyperref}

\begin{document}
\title{FliX: Flipped-Indexing for Scalable GPU Queries and Updates}

\author{
\IEEEauthorblockN{Rosina Kharal}
\IEEEauthorblockA{University of Waterloo\\ Waterloo, Canada\\ rosina.kharal@uwaterloo.ca}
\and
\IEEEauthorblockN{Trevor Brown}
\IEEEauthorblockA{University of Waterloo\\ Waterloo, Canada\\ trevor.brown@uwaterloo.ca}
\and
\IEEEauthorblockN{Justus Henneberg}
\IEEEauthorblockA{Johannes Gutenberg University\\ Mainz, Germany\\ henneberg@uni-mainz.de}
\and
\IEEEauthorblockN{Felix Schuhknecht}
\IEEEauthorblockA{Johannes Gutenberg University\\ Mainz, Germany\\ schuhknecht@uni-mainz.de}
}

\maketitle

\begin{abstract}

GPU concurrent data structures (CDSs) achieve very high read-only throughput, but fully dynamic updates on GPU-resident data remain challenging. Ordered CDSs (e.g., B-trees and LSM-trees) route operations through an \emph{index layer} to a \emph{data layer} of buckets. Maintaining and traversing that index under updates causes contention, warp divergence, and memory overhead. Hash tables avoid this cost but lose range and successor queries.

We present \flix{}, a comparison-based \emph{flipped} indexing strategy that eliminates the index layer. Instead of assigning a warp to each operation, \flix{} assigns a warp to each \textit{bucket}. Each bucket then binary-searches a sorted batch for its operations, so one search replaces many index traversals and updates become simple. \flix{} reduces query latency by up to $6.5\times$ versus a leading GPU \btree{} and $1.5\times$ versus the leading GPU \lsmtree{}. It delivers $4\times$ higher \textit{throughput per memory footprint} than ordered competitors and over $8\times$ higher insertion throughput than the closest fully dynamic ordered baseline. It also matches or beats unordered GPU hash tables on point queries and deletions, so its ordering support comes at virtually no cost.

\end{abstract}

\section{Introduction}
\label{sec:introduction}

\begin{figure*}[t]
  \centering

  \includegraphics[
    page=7,
    width=0.95\textwidth,
    trim={0cm 33cm 1cm 0.0cm},
    clip
  ]{Figures/gpu_updates_conceptual_figures_latest.pdf}
  \vspace{-2mm}
  \caption{ \small (a) Index layer search required for each query key. (b) Reduce index layer searches if bucket ranges are known in advance. (c) Eliminate the index layer by performing binary search on the sorted batch of queries.}
  \label{fig:flix-conceptual}
   \vspace{-2mm}
  \Description{Conceptual Design of Flix}

\end{figure*}

Recently, GPUs have become increasingly attractive as processing accelerators in hybrid CPU-GPU query engines~\cite{lit:arch1, lit:boss, lit:maximus, lit:sirius, lit:gpuinterconnects}, where heavy parts of analytical queries, such as joins or group-by operations, can be offloaded on demand to the GPU.
However, as real-world workloads are often hybrid and consist not only of read-only analytical queries, but also of modifying transactions (so-called HTAP workloads), this raises the question of how to ensure that analytics on the GPU operate on fresh data.
A naive way of ensuring freshness is to ship a current snapshot of the data over to the GPU every time an operation is offloaded. However, this heavily penalizes the CPU-GPU interface. A more elaborate way is to keep data cached in GPU memory and to maintain its freshness, e.g., by periodically shipping a delta over to the GPU and applying it as a batch operation in one go~\cite{lit:caldera, lit:gputestoftime}.

Of course, if data is cached on the GPU, then not only base columns must be kept fresh, but also any auxiliary index structures referencing these columns or intermediate results from previously processed operations. Unfortunately, the current situation is that index structures on the GPU are typically not maintained, but fully rebuilt~\cite{lit:rtindex, lit:cgrxu, lit:hash-cudpp1, lit:learnedindex}. There is a simple reason for this: While GPU-based index structures are able to achieve extremely high-speed read-only query operations, often vastly outperforming CPU-based data structures~\cite{lit:btree1, lit:gpuskiplist, lit:hash-slabhash, lit:warpcore, lit:mitgpucpu}, they either completely lack support for updates or handle them poorly.
For example, the GPU \lsmtree{} of Ashkiani et~al.~\cite{lit:lsmtree} supports insertions and deletions at the cost of very high memory consumption.
The ray-tracing powered \textbf{RTIndeX} of Henneberg et~al.~\cite{lit:rtindex}
and the cuDPP static hash table~\cite{lit:hash-cudpp1} require a full rebuild of the index on insertions and deletions.
The Warpcore hash table \warphash{} of J{\"{u}}nger et~al.~\cite{lit:warpcore} supports insertion and deletion, but the table has a fixed size and cannot grow or shrink.
The GPU~\btree{} of Awad et~al.~\cite{lit:btree1} supports both insertion and deletion, but as our experiments reveal, their approach to supporting fully dynamic updates incurs high overhead. Some of these also have limited or no support for reclaiming memory.
The reasons for these limited update capabilities are rooted in the fact that it is notoriously difficult to efficiently perform highly concurrent structural modifications which are required to fully utilize the GPU. In traditional index structures, which consist both of an \textit{index layer} and a \textit{data layer} (e.g., the inner nodes and the leaf nodes of a B-tree), this becomes visible: The index layer is a global structure over the data, which carefully needs to be modified by a large number of threads while ensuring correctness. \textit{But what if we could eliminate the index layer altogether? Would this maybe not only simplify updates, but also usage in general?}

\subsection{Flipping the Index --- \flix{}}
To understand the key design of~\flix{}, consider the following example involving the~\btree{} of Awad et al.~\cite{lit:btree1}. For simplicity, we do not start with updates, but with the case of answering a batch of point queries, as they might be requested by a database join.
For the \btree{}, the work is divided as follows:
Each GPU warp takes 32 consecutive queries from the batch, and the warp then answers these queries, one by one, collaboratively.
In other words, for each warp, for each key, the warp traverses the \btree{}'s index layer to reach the correct bucket in the data layer (Figure~\ref{fig:flix-conceptual} (a)).
The reason for having an entire warp collaborate on a single index lookup is that performing simultaneous lookups of different keys would likely cause threads to traverse different paths from one another, causing \textit{warp divergence}, which occurs when threads in a warp take different branches.
One could induce temporal locality by first sorting queries by their keys, so that the queries a warp is responsible for are more likely to be located close to one another in the tree.
This way, consecutive queries for nearby keys can benefit from nodes being cached by prior queries.
However, depending on the workload, query keys may simply be too far apart for threads to realize caching benefits.
Even if many keys are close enough to realize benefits, a lot of time could be spent performing redundant searches, one by one, through the same parts of the tree.

What if we could guarantee that each warp were \textit{always} assigned keys destined for a \textit{single bucket}? Then we could simply perform a \textit{single traversal} through the index layer.
But, of course, a warp does not know which keys are destined for a given bucket until it has traversed the tree and reached the bucket.
Imagine, then, that we flip this problem on its head and ask the opposite question:
Rather than trying to determine \textit{which bucket a query key is destined for},
we begin at the buckets,
and ask \textit{which keys belong in this bucket?}
In this \textit{flipped} paradigm, instead of traversing the index layer to map a key to a bucket, we traverse the query batch instead, to map a bucket to a set of queries, completely eliminating redundant traversals.
If each bucket records which part of the key range it is responsible for, one could build an index over the incoming queries, and use that index to search for the smallest and largest key that could appear in a given bucket (Figure~\ref{fig:flix-conceptual} (b)).
In fact, we can eliminate the index layer and traversals altogether, by simply sorting the query list (which is extremely fast on the GPU) and performing a single binary search to obtain the queries destined for a bucket (Figure~\ref{fig:flix-conceptual} (c)).
With this approach, we also avoid having to structurally modify an index layer under updates and instead limit all modifications to the data layer, that is, to our buckets.

This approach leverages the extreme scale parallelism of GPUs to do exactly what GPUs are good at.
Each bucket is effectively a task to be performed in a GPU kernel.
In this kernel, a warp first rapidly determines whether the bucket it is responsible for has any corresponding operations in the batch, be it a query or an update, and if not, it terminates.
Any bucket that survives this first step, then performs the operations destined for it sequentially.
In essence, whereas traditional GPU indexes map \textit{compute onto operations}, which then locate buckets, \flix{} maps \textit{compute onto buckets}, which then locate relevant operations.

The approach also raises three challenges.
First, a B-tree balances data automatically, whereas \flix{} must partition keys into buckets and keep that partitioning balanced as the distribution shifts.
Second, because compute is assigned to buckets, skew in a batch could translate directly into skew in the work done per bucket.
Third, \flix{} requires each batch to be sorted.
Section~\ref{sec:paradigm} shows that each has a surprisingly simple answer: restructuring rebalances buckets entirely on the GPU, the default mapping proves robust to all but extreme skew, and GPU sorting is cheap enough that \flix{} still outperforms every baseline when it is charged for sorting unsorted batches.

\subsection{Contributions}

\noindent \textbf{(1)}~We present \flix{}, a novel \textit{flipped indexing} paradigm and GPU CDS with dynamic updates and high-performance queries that is well suited to database indexing.
Unlike prior art, \flix{} starts at a \textit{bucket} in the data layer and searches for the operations targeting that bucket, allowing all updates to a bucket to be performed together with coalesced accesses.

\noindent \textbf{(2)}~We exploit the order maintained by \flix{} with range query kernels that
traverse wide ranges in parallel, and with a range aggregation variant whose cost is independent of range size.

\noindent \textbf{(3)}~We extend \flix{} to fully concurrent execution of mixed batches of insertions, deletions, and queries in a single kernel launch, with a guaranteed order of operations on each key.

\noindent \textbf{(4)}~We propose GPU resident restructuring for \flix{} to maintain low query latencies in long-running applications with many insertions and deletions, and to bound memory consumption by gradually merging \textit{underfull} nodes (nodes with few keys).

\noindent \textbf{(5)}~We conduct \textit{formative} synthetic experiments to identify the most effective GPU kernel configurations across different update/query ratios, key distributions, and batch sizes.
This exploration is followed by extensive \textit{summative} experiments over a wide variety of workloads that demonstrate \flix{}'s strengths against the state of the art.

\begin{figure}[h!]
  \centering
  \includegraphics[width=0.98\columnwidth]{
    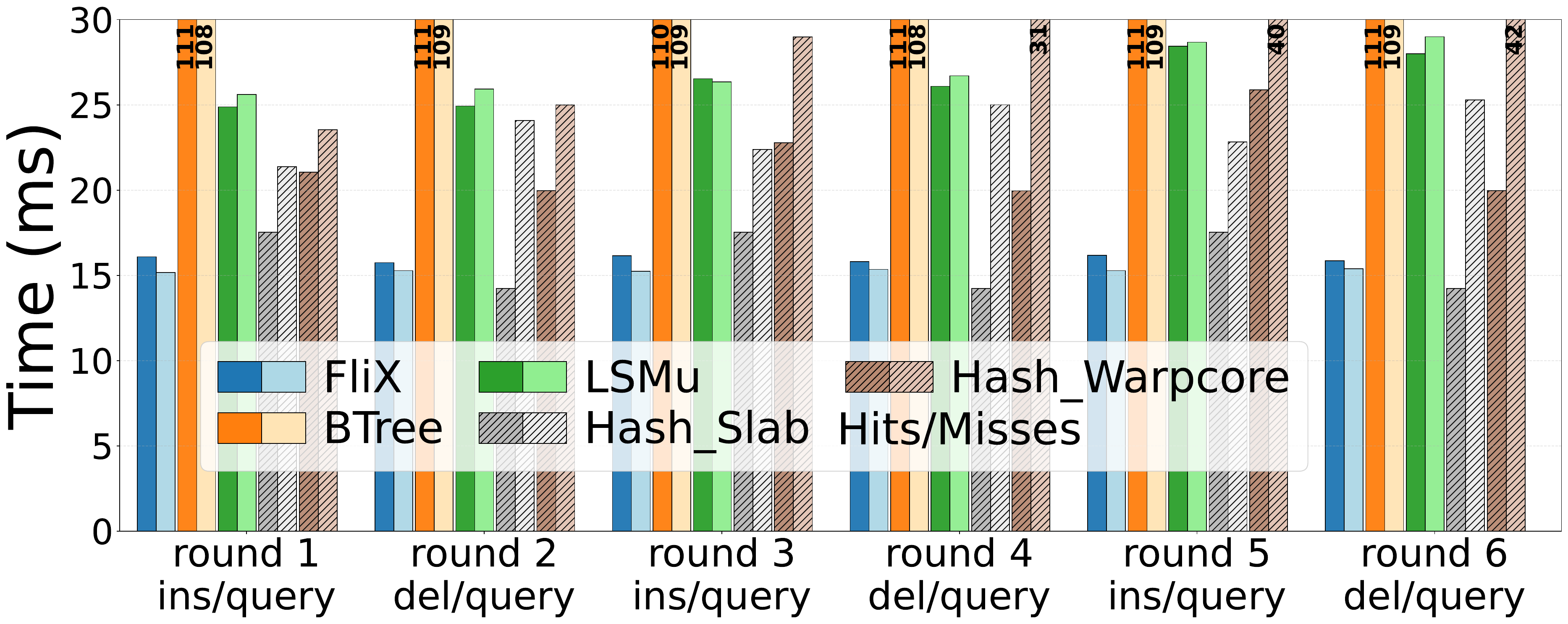
  }

  \caption{ \small State-of-the-art GPU CDSs against \flix{}; alternating rounds of updates, illustrating query latency (ms) after each update. For each structure, the darker bar shows a batch of \textit{all-hit} queries (keys present in the structure) and the lighter bar shows a batch of \textit{all-miss} queries (keys not present).}

  \label{fig:teaser01}
  \Description{Query latency comparison of GPU data structures.}
\end{figure}

\subsection{Key Results}
Figure~\ref{fig:teaser01} previews our experiments: \flix{} is competitive with the state of the art on point queries that all hit, and outperforms it when they all miss.
Strikingly, two of the indexes \flix{} outperforms are \textit{hash tables}, which pay nothing to maintain order.
We include them not as competitors but to measure the price of order, which our results show is marginal.

Against its ordered competitors, \flix{} answers point queries up to $6.5\times$ faster than the GPU \btree{} and $1.5\times$ faster than \lsmu{}, and delivers over $4\times$ their query throughput per byte of memory.
It inserts $8\times$ faster than the \btree{}, and deletes faster than every baseline except \lsmu{}, whose speed comes at $2.8\times$ \flix{}'s memory footprint.
Order then pays off directly: across range sizes, \flix{} answers range queries $1.4\times$ and $5.0\times$ faster than \lsmu{} and the \btree{}, and range aggregations $2.9\times$ and $14.9\times$ faster (geometric means).
Finally, in fully concurrent execution,~\flix{} performs batches of mixed updates and queries at $4.42\times$ higher throughput than~\slabhash{} and $11.39\times$ higher throughput than the~\btree{}.

\section{Background and Related Work}
\label{sec:motivation_related}
GPU CDSs typically process operations in batches: operations are collected on the host, copied to the GPU as a batch, processed by a GPU kernel, and their results are copied back.
A batch usually contains a single operation type; some data structures also support batches of mixed operation types, which we discuss in Section~\ref{sec:baselines-fully-concurrent}. In Section~\ref{sec:flix-fully-concurrent}, we show how \flix{} handles mixed operations.

\subsection{\mbox{Concurrent Data Structures and Updatability}}

CDSs used for database indexing typically implement an unordered or ordered map from keys to rows, supporting \textit{search}, \textit{insert}, and \textit{delete}.
Ordered maps additionally support \textit{range queries}, which return all keys (and values) in a key range, and \textit{successor} queries, which return the next larger key.

Some GPU CDSs support updates only through a \textit{full rebuild}, reconstructing the structure from the live keys and the new batch~\cite{lit:hashfightlessley2020, lit:rtindex, lit:shahvarani2016hybrid}.
Others allow fine-grained updates but require periodic \textit{partial rebuilds} to keep access efficient~\cite{lit:lsmtree}.
We say a GPU CDS offers \textit{dynamic updatability} if it supports insertions and deletions during long-running execution without either.
To be practical, it should also reclaim memory, so that long-running applications do not grow without bound.

Since queries dominate most indexing workloads, an updatable GPU CDS should still prioritize query performance.
State-of-the-art GPU structures answer read-only queries quickly~\cite{lit:hash-cudpp1, lit:hash-slabhash, lit:warpcore, lit:btree1, lit:fastinmemory}, but, as our experiments show, they do not always scale under frequent updates.
Updates are harder because tens of thousands of threads must be coordinated in a SIMT (Single Instruction, Multiple Threads) architecture, and because peak performance requires coalesced memory accesses. Updates that are scattered across an index break these coalesced memory accesses~\cite{lit:awaddissertation, lit:lsmtree}.
Such scattered accesses are precisely what motivates our flipped indexing strategy.

\begin{figure*}[!t]

  \centering
  \includegraphics[
    page=2,
    width=0.9\textwidth,
    trim={0cm 32cm 5cm 0cm},
    clip
  ]{Figures/gpu_updates_conceptual_figures_latest.pdf}
  \vspace{-2mm}
  \caption{\small (a) \flix{} initial build step: keys are divided into buckets. (b) A batch of keys is inserted.
  Each thread fetches a key from the current node ([8, 11, -, -]), and compares insert keys to its fetched key to determine where new keys go.
  (c) A batch of keys is deleted.
  (d) Restructuring occurs, flattening chains of nodes into individual buckets.
  Underfull nodes within the same bucket are merged to reclaim space.
  }
  \label{fig:flix-combo-rebuild}
  \Description{flix initial build, insertions, deletions and including rebuilding}
\end{figure*}
\subsection{Updatable GPU Data Structures}
\label{sec:updateable-data-structures}

\textbf{GPU LSM-Tree}.
The~\lsmtree{}~\cite{lit:lsmtree} of Ashkiani et al.\ adapts the \textit{Log-Structured Merge} tree~\cite{lit:traditionalLSM} to the GPU.
Insertions are batched into sorted chunks that are merged into levels of geometrically increasing size, which exploits fast GPU sorting and merging and yields excellent insertion throughput.
However, merges are not performed in place and require auxiliary buffers proportional to the largest level, so memory overhead is high, and a query must search every level.
Deletions insert tombstones, which, like stale keys, persist until a cleanup step compacts the structure, so a long-running GPU-resident \lsmtree{} requires repeated cleanup.
Our experiments use \lsmu{}, an improved variant described in Section~\ref{sec:baselines-for-experiments}.

\textbf{GPU B-Tree}.
The~\btree{} by Awad et al.~\cite{lit:btree1} introduces a dynamic, concurrent B-Tree optimized for warp cooperative execution on GPU architectures. It can support point, range, and successor queries as well as update operations, providing a fully mutable GPU-resident ordered index. \btree{} nodes are aligned to 128-byte cache lines and organized as linked chains through side-link pointers following B-Link tree principles, allowing safe concurrent traversals. A warp-cooperative work sharing (WCWS) strategy minimizes divergence, and proactive splitting of nodes with restart-on-failure policies reduces contention during insertions.
The~\btree{} is the closest comparable ordered CDS to~\flix{}, supporting all the same operations and compacting space immediately on deletions.

\textbf{GPU Hash Tables}.
The slab-based hash table \slabhash{} of Ashkiani et al.~\cite{lit:hash-slabhash} stores \textit{key-value} pairs in linked chains of fixed-size slabs, allocated on the GPU by a custom allocator without CPU intervention.
Deletions are logical at first; physical removal is deferred to a compaction phase.
\warphash{}~\cite{lit:warpcore} uses open addressing with warp-cooperative probing over a pre-allocated table of fixed size, so its memory use is set by the table size and load factor rather than by the number of keys.
Its deletions leave tombstones, which are never reclaimed but can be reused by later insertions.
Both hash tables achieve state-of-the-art update throughput, but, being unordered, cannot efficiently support range or successor queries.

\textbf{Fully Concurrent Execution}.
\label{sec:baselines-fully-concurrent}
The~\btree{} by Awad et al.~\cite{lit:btree1} and~\slabhash{} by Ashkiani et al.~\cite{lit:hash-slabhash} support batches of \textit{mixed} operation types on the GPU with fully concurrent execution.
This is difficult to do efficiently due to the large number of threads on a GPU.
It also complicates experimental correctness checks, since the outcome depends on the ordering of competing inserts and deletes.
This raises an interesting question:
What order \textit{should} conflicting operations be performed in?
The~\btree{} serves operations within a warp in input order.
\slabhash{} processes operations within a warp in priority order starting with insertions, then deletions, and finally queries.
In both, ordering across warps is a function of the scheduler.
\flix{} takes a similar approach to \slabhash{}, but we also order operations across an entire batch, not just within individual buckets.

\section{The \flix{} approach}
\label{sec:paradigm}

\subsection{Compute-to-Bucket Mapping}
\label{sec:flixbackground}
GPU indexes have traditionally mapped \textit{compute to operations}: each thread or warp takes an operation, such as a query for key $k$, and locates the bucket it needs.
\flix{} instead uses a \textit{compute-to-bucket} mapping, in which each thread or tile is responsible for a single bucket and pulls its work from the batch of operations.
A \textit{tile} is a group of threads within a warp that cooperate on a task, expressed with cooperative groups~\cite{lit:coopgroupsharris2017}; a 32-thread warp can be split into tiles of 16, 8, or fewer threads.
Existing GPU CDSs~\cite{lit:gpuskiplist, lit:lsmtree, lit:btree1, lit:hash-cudpp1, lit:hash-megakv, lit:hash-slabhash, lit:warpcore}, including those in Section~\ref{sec:motivation_related}, follow the compute-to-operation model.
Their layouts offer no fixed, contiguous unit at which compute could be stationed, so adapting them to this model is non-trivial.

\subsection{Data Structures \& Initial Build}
\label{sec:flixBuild}

Figure \ref{fig:flix-combo-rebuild} (a) provides a visual illustration of the initial build stage of \flix{} based on an initial set of \textit{key-rowID} pairs.
The build keys are sorted and grouped into partitions of size \textit{p}, and the groups determine which keys belong in each bucket.
Values (\textit{rowIDs}) corresponding to each key are left out of the diagram for simplicity.
The capacity of each node, $nodesize$, is 4 in the example, and we set $p=\frac{nodesize}{2}$ so that nodes are initially \textit{half} full.
The initial fill state is a tunable parameter.
Less full nodes will support more insertions before node splitting is required.
The largest key per group (every $p^{\text{\tiny th}}$ key) is the maximum allowable key for the corresponding bucket (\texttt{maxBucket}).
The maximum allowable keys for each bucket are stored in an array called the
\textit{max key per
bucket array} (\textbf{MKBA}), which is omitted from our diagrams for simplicity.

At build time, each bucket consists of a single node. As additional keys are inserted into \flix{}, new linked nodes may be added to a bucket; therefore, a bucket is formed by a chain of one or more nodes using \texttt{node-link} pointers. The \texttt{maxBucket} values are used to distinguish ranges across buckets (\textit{inter-bucket} ranges), while node-level maximums, called \texttt{maxKey}, distinguish ranges within a bucket (\textit{intra-bucket} ranges). Each node stores two pieces of metadata: its \texttt{maxKey} and size (the number of keys currently stored). For simplicity, the size field is omitted from the figure.

Figure \ref{fig:flix-combo-rebuild} (b) illustrates the insertion process after the initial build, where a tile of 4 threads processes keys from the insert list.
At a high level, the tile advances key by key in the insert list, and for each key $k$, the threads in the tile collaborate (by taking ownership of a key in the node to assist in comparisons) to determine the correct location in the node where the key should reside.
If a key is not already present, it is inserted into the node.
If the node is full, it is split into two halves, and the insertion continues in the appropriate node.
When a split occurs, the \texttt{maxKey} values are updated to reflect the new key ranges of the two nodes.
Although node-level maximums may change over time, the \texttt{maxBucket} values remain fixed unless restructuring occurs
(Section \ref{sec:rebuilddescription}).
A more detailed description of the algorithm appears in Section~\ref{sec:algorithmcondenseddescriptions}.

Figure~\ref{fig:flix-combo-rebuild} (c) illustrates the state of the data structure following a series of deletions. In \flix{}, deleted keys are \textit{physically removed} immediately, with surviving keys shifted to compact the node and reclaim the free space. This is in contrast to prior GPU CDSs, which commonly perform deletions through the use of tombstones; whereby a key is logically deleted but remains physically resident until a deferred cleanup phase. This approach allows logically deleted entries to accumulate over time, effectively introducing garbage into the data structure. It reduces usable memory and degrades query efficiency, since queries must explicitly identify and skip tombstoned entries.
The~\lsmtree{},~\slabhash, and~\warphash{} all employ tombstone-based deletions.

\begin{figure}[t]
   \centering

  \includegraphics[
    page=8,
    width=0.8\linewidth,
    trim={3cm 38cm 60cm 0cm},
    clip
  ]
  {Figures/gpu_updates_conceptual_figures_latest.pdf}

  \caption{ \small In \flix{}, each bucket binary searches the query list and \textit{pulls} relevant keys.
  }
  \vspace{-2mm}
  \label{fig:paradigm-shift1}
  \Description{Query workload- visualization}

\end{figure}

\begin{table}[h!]
\centering
\captionsetup{font=small,skip=2pt}
\setlength{\abovecaptionskip}{2pt}
\setlength{\belowcaptionskip}{0pt}
\renewcommand{\arraystretch}{1.0}
\setlength{\tabcolsep}{6pt}
{\small
\begin{tabular}{cc|cc}
\hline
\textbf{Batch size} & \textbf{Sort time} & \textbf{Build size} & \textbf{Sort time} \\
\hline
$2^{18}$ & 0.056 & $2^{24}$ & 1.656 \\
$2^{20}$ & 0.130 & $2^{26}$ & 6.513 \\
$2^{22}$ & 0.442 & $2^{28}$ & 25.991 \\
\hline
\end{tabular}
}
\caption{\small Average Sort time (ms) on NVIDIA A6000 GPU}
\vspace{-2mm}
\Description{Table of query times on GPU A6000}

\label{tab:avg_sort_time_a6000}
\end{table}

\subsection{Point Queries}
\label{sec:queries}

Figure~\ref{fig:paradigm-shift1} illustrates the basic constructs of our data structure, which presupposes the query list is sorted. Sorting and merging are operations GPUs can perform very efficiently; the overhead costs of sorting the query list are discussed further in Section \ref{sec:query-experiments} and Table \ref{tab:avg_sort_time_a6000}.
Compute units stationed at each bucket pull relevant operations by performing a simple binary search on the query list, thereby taking all operations belonging to the bucket and executing them in place. As shown in Figure~\ref{fig:paradigm-shift1}, each bucket will pull its relevant portion of query keys from the main query list. This effectively removes the need for a centralized index to direct traffic.
As explained above, the \texttt{maxBucket} value determines the maximum allowable key supported by bucket $b_i$. A single binary search for the key \texttt{maxBucket$_{b_{i-1}}$} is sufficient to retrieve the starting index of query keys belonging to \textbf{$b_i$}; we continue to perform queries in $b_i$ from this index until a key exceeds \texttt{maxBucket$_{b_{i}}$}. Queries return a \textit{result buffer} to the host for the corresponding values (\textit{rowIDs}) associated with each search key, or a \textit{not found} value.

\newcommand{\rqbucket}{\textsc{RQ-Bucket}}
\newcommand{\rqquery}{\textsc{RQ-Query}}
\newcommand{\rqspan}{\textsc{RQ-Span}}

\subsection{Range Queries (RQs)}
\label{sec:range-queries}

An RQ retrieves the values (or rowIDs) of all keys between a lower and an upper bound key.
Depending on the application, the result is either the array of these values or a compact digest.
An RQ covers a contiguous run of buckets, since they are ordered.
We present a basic kernel and two optimizations.

\textbf{Basic kernel (\rqbucket).}
Each bucket takes responsibility for RQs whose lower bounds fall in its range, and processes them one at a time.
For each RQ, the bucket traverses consecutive nodes, continuing into following buckets as needed, until it sees a \texttt{maxKey} that ends the range.
At each node, one thread reads the node metadata and shares it with the rest of the tile, then every thread checks one key against the range, and the tile records the qualifying values.

\textbf{Assigning work by query (\rqquery).}
\rqbucket{} works well for small queries, but if 100 massive queries have low endpoints in the same bucket, that bucket is responsible for them.
For queries that span many buckets, compute-to-query is a better model.
\rqquery{} assigns each query to a \textit{thread}, packing consecutive (sorted) queries into tiles.
The tile then serves its threads' queries one at a time, all threads cooperating on each.
Every thread reads node metadata instead of waiting for one thread to share it, and collects its own part of the result; the tile combines these parts once per query (vs once per node).
Tiles may be half a warp wide, letting a warp run two queries at once.

\textbf{Parallelizing wide ranges (\rqspan).}
\rqquery{}'s flaw is that a query traverses one node at a time.
\rqspan{} first locates the lower and upper bound buckets, processing them as in \rqquery{}.
The buckets in between are then distributed over the threads of the tile; threads do not need to read or compare their keys.
Crucially, each bucket $b$'s first node is at a predictable address, $\mathit{base} + b \cdot \mathit{nodeSize}$, so all threads can start at the same time.
A tile thus keeps \textit{one traversal in flight per thread}.
Binary searching for the end bucket adds overhead to every query.
\rqspan{} mitigates this cost with \textit{galloping search}.
Queries are sorted, so consecutive queries served by a tile often start nearby.
Each search therefore starts at the previous query's bucket, probes forward in steps that double in length (exponential search), and finishes with a binary search.

\textbf{Alternative aggregation variant.}
We also implement a variant that precomputes prefix aggregates of an invertible aggregation function $\Phi$~\cite{ho1997range} (e.g., sum, count, or XOR, and metrics derived from them such as mean and variance) to facilitate fast $\Phi$ computation over a batch of ranges.

\subsection{Distributional Shift}

\label{sec:distributional-shift-description}
A distributional shift occurs when operations concentrate in a small region of the key space.
For example, after a uniform build, insertions and queries may come to target only 5\% of the bucket range.
Insertions then create long node chains in these \textit{hot} buckets, and queries spend longer traversing them, so under the default \textit{one-unit-of-compute-per-bucket} mapping the tiles assigned to hot buckets may become overloaded.

\flix{} can mitigate this by changing how compute is assigned to buckets.
An \textit{adaptive} strategy assigns more or fewer threads to a bucket as needed, while an \textit{elastic} strategy splits buckets with long chains, reducing the key range each one covers.
In practice, the default mapping proves surprisingly robust even under highly skewed workloads (Section~\ref{sec:highlyskewed-queries}).
An example of an adaptive strategy is our \rqquery{} range query kernel (Section~\ref{sec:range-queries}), which
assigns compute per query rather than per bucket.
The next subsection describes an elastic strategy.

\subsection{Restructuring}
\label{sec:rebuilddescription}

To address both distributional shift and sustained growth of the index, we provide a \textit{restructuring} procedure that flattens large buckets containing many nodes into individual buckets (with one node per bucket).
This is illustrated in Figure \ref{fig:flix-combo-rebuild} (d).
This procedure effectively realigns the buckets (and compute) to the current distribution such that keys map uniformly to buckets, and can improve query performance by eliminating traversals of long chains of nodes.
If there is no distributional shift, and the size of the index is relatively stable, then restructuring may be unnecessary.
One can think of the frequency of restructuring as a tunable knob that allows a system to invest time in maintaining the structure.
Restructuring also reclaims memory by merging \textit{underfull} nodes.
In Section~\ref{sec:restructuring-experiments}, we quantify impact on reclaimed memory and query performance.

\begin{table}[t]
\centering
\captionsetup[table]{font=small}
\renewcommand{\arraystretch}{0.8}
\setlength{\tabcolsep}{6pt}

{\small
\begin{tabular}{@{}L{2.6cm}|cccccccc@{}}
\toprule

$curr$ &
\green{10} & \green{25} & \green{30} & \green{40} &
\green{70} &  &  &  \\
\midrule

Thread $t_i$        & 0 & 1 & 2 & 3 & 4 & 5 & 6 & 7 \\
key in $t_i$ regs   &
\green{10} & \green{25} & \green{30} & \green{40} &
\green{70} &  &  &  \\
\midrule

Insert $sublist_i$ &
 & \textbf{15} & \textbf{17} & \textbf{39} & \fbox{\textbf{65}} &  &  &  \\
\midrule

{\scriptsize step 1: \textit{test-key}=25, $curr$} &
\green{10} & \textbf{15} & \textbf{17} & \green{40} & \green{70} & & & \\
{\scriptsize step 2: \textit{test-key}=40, $curr$} &
\green{10} & \textbf{15} & \textbf{17} & \green{25} & \green{30} & \textbf{39} &  & \\
{\scriptsize step 3: \textit{test-key}=70, $curr$} &
\green{10} & \textbf{15} & \textbf{17} & \green{25} & \green{30} & \textbf{39} & \green{40} & \green{70} \\

\bottomrule
\end{tabular}

}
\caption{\small TL-Bulk in-place insertions. Threads lift original keys from $curr$ into registers (\green{green}) and merge $sublist_i$ in place using successor boundaries (\textit{test-keys}).
Key $65$ from $sublist_i$ will be inserted following a node split of $curr$.}
\label{tab:TL_bulk_insert_example}

\end{table}

\section{Update Algorithms}
\label{sec:algorithmcondenseddescriptions}

Our primary design goal in~\flix{} is to offer query performance that is competitive with the state-of-the-art, and subject to that constraint, to offer high performance for updates.

Modern GPUs provide different approaches for orchestrating the work done by a warp, and these approaches naturally lead to different update algorithms.
We explore and compare two broad approaches previously used in the literature: single-threaded (ST) kernels~\cite{lit:rtindex,lit:lsmtree}, and tiled (TL) kernels~\cite{lit:btree1,lit:hash-slabhash, lit:warpcore}. While previous work may choose to select one approach over the other, we perform a comparative analysis of the various types of kernels to evaluate the applicability of each.

ST kernels are expressed as a single thread of execution, such that threads in a warp are independently executing tasks.
In~\flix{}, each thread operates on its own bucket, and insofar as the hardware can support doing so, multiple threads can run in parallel on the same SM, accessing mutually exclusive buckets.
TL kernels, on the other hand, involve multiple threads in a warp collaborating on a single bucket.
Recent hardware with support for cooperative groups can schedule multiple tiled subgroups to execute concurrently on a single warp~\cite{lit:coopgroupsharris2017}.
ST kernels are simpler, but can suffer greater thread scheduling overhead and warp divergence compared to TL.

The specific update algorithm at the bucket level may vary; however, the preprocessing steps for queries, insertions and deletions are common across all algorithms.
We refer to a batch of keys to be inserted or deleted as an \textit{update list}, and a batch of keys to be queried as a \textit{query list}.

\subsection{Common Steps}
\label{sec:Alg-commonsteps}
The \flix{} approach assigns one \textit{unit of compute} to each bucket. A unit of compute is a single thread in an ST algorithm, and a tile in a TL algorithm. Suppose there are $n$ buckets. \flix{} then assigns $n$ units of compute, $c_1, ..., c_n$, to buckets, $b_1, ..., b_n$, respectively.

In a similar manner to how queries are performed using binary search on the query list (Figure \ref{fig:paradigm-shift1}), each~\texttt{$c_i$} retrieves a $sublist_i$ containing the keys destined for bucket $b_i$ from the \textit{update list}.
For a given key~$k$ from $sublist_i$, we traverse the nodes in $b_i$ by following \texttt{node-link} pointers until we reach the node where $k$ belongs --
that is, the first node whose \texttt{maxKey} satisfies $k \leq maxKey$.
In the algorithms that follow, the current node is denoted by $curr$.
At this point, the update operation proceeds at the bucket level.

\begin{table}[t]

\centering
\captionsetup[table]{font=small}
\renewcommand{\arraystretch}{0.85}
\setlength{\tabcolsep}{6pt}
\small
\begin{tabular}{@{}L{3cm}|cccccccc@{}}
\toprule

$curr$ &
\green{10} & \green{15} & \green{20} & \green{25} &
\green{30} & \green{35} & \green{40} & \green{45} \\
\midrule

Thread $t_i$      & 0  & 1  & 2  & 3  & 4  & 5  & 6  & 7 \\
key in $t_i$ regs &
\green{10} & \green{15} & \green{20} & \green{25} &
\green{30} & \green{35} & \green{40} & \green{45} \\
\midrule

Delete list       & \textbf{20} & \textbf{30} &  &    &  &    &    &    \\
Tile Mask (deleted)    & 0  & 0  & 1  & 0  & 1  & 0  & 0  & 0 \\
\midrule

Shift-left distance    & 0  & 0  & -- & 1  & -- & 2  & 2  & 2 \\
Compacted keys    & \green{10} & \green{15} &    & \green{25} &    & \green{35} & \green{40} & \green{45} \\
\midrule

final state $curr$      &
\green{10} & \green{15} & \green{25} & \green{35} &
\green{40} & \green{45} &  & \\
\bottomrule
\end{tabular}

\caption{ \small TL-Bulk deletions with compaction. Each thread $t_i$ loads one key from $curr$. Keys 20 and 30 are removed; remaining keys shift left by the number of prior deletions.}
\label{tab:TL_bulk_delete}

\end{table}

\subsection{Single Threaded Insertions}
\label{sec:STAlgorithms}
\textbf{Shift-Right Insertions} (\texttt{ST-Shift-Right}).
The current node, $curr$, is searched using binary search to check for the existence of key $k$ from $sublist_i$. If not present, the appropriate insertion point in $curr$ is returned. Keys ahead of the insertion point are shifted right, and the new \textit{key-rowID} pair is inserted. If $curr$ is determined to be full, a \texttt{node\_split} function is called to split the keys into two halves. Subsequently, the insertion process continues at the appropriate node.

\textbf{Bulk Insertions} (\texttt{ST-Bulk}).
A single thread bulk-inserts all keys from $sublist_i$ into $b_i$ using a copy space. The thread merges $sublist_i$ and the contents of $curr$, copies them back to the original node and handles any necessary node splitting within the copy-back step. Thread-local memory was the most performant strategy for maintaining a node copy space.

\begin{figure*}[t]
    \centering
    \begin{tabular}{c}
        \includegraphics[width=18cm]{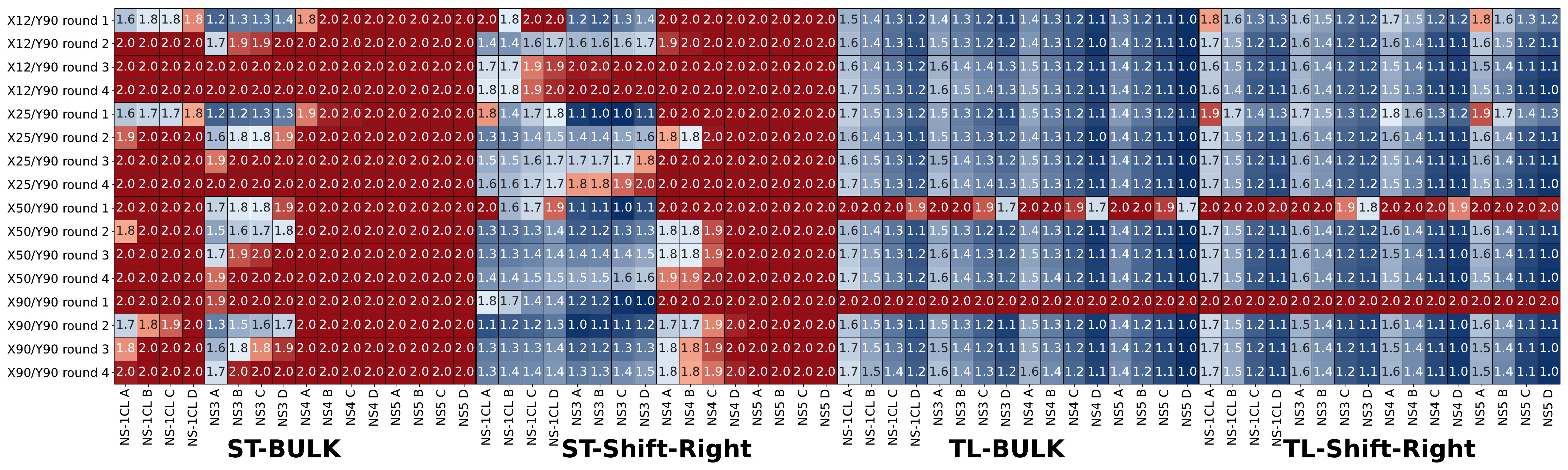} \\
    \end{tabular}
    \vspace{-2mm}
    \caption{ \small Full heat map across varying insertion workloads with 4 rounds of insertions.
    The x-axis represents insertion algorithms, each evaluated under 16 configurations (4 $nodeSize$ configurations $\times$ 4 $threadsPerBlock$ assignments (A-D)). Results indicate that \textit{Tiled Bulk} insertions outperform alternatives in all rounds except the first round for more uniform workloads ($X=50$, $Y=90$ \& $X=90$, $Y=90$).}
    \label{fig:full-heatmaps}
    \Description{Heat map of insertions in FliX}

\end{figure*}

\subsection{Cooperative Groups: Tiled Insertions}
\label{sec:tiledalgorithms}
\textbf{Tiled Shift-Right Insertions} (\texttt{TL-Shift-Right}).
Similar to ST, but performed in parallel by all threads in a tile. Each thread reads one key from $curr$; shift-right is performed using parallel comparisons and writes.

\textbf{Tiled Bulk Insertions} (\texttt{TL-Bulk}).
Threads in $tile_i$ are tasked with inserting keys from $sublist_i$ into bucket $b_i$ at the appropriate node.
We avoid a copy space by lifting all original keys from a node into thread-private storage (typically registers). The TL-Bulk algorithm is illustrated with an example in Table~\ref{tab:TL_bulk_insert_example}. The steps are:

{\small
\noindent (1)~\textbf{Load node state:}  Threads in $tile_i$ each load (\textit{lift}) one key from $curr$. These keys are denoted the \textit{original keys}.

\noindent  (2)~\textbf{Compute successor boundary:} For the current insert key, $k$ in $sublist_i$, threads in $tile_i$ collectively determine the \textit{successor key} in $curr$, denoted \textit{test-key} (the smallest original key $>k$). If no such key exists, \textit{test-key} is $+\infty$.

\noindent  (3)~\textbf{In-place merge for one boundary:} For all insert keys $k'$ where $k\le k'<\textit{test-key}$, one thread from $tile_i$ writes $k'$ into the next free position in $curr$, overwriting original keys in $curr$.

\noindent (4)~\textbf{Write-back original keys:} When the next $k'$ reaches or exceeds \textit{test-key}, compute \textit{test-key'}, the next original key $> k'$. Threads write back all original keys in the interval [$\textit{test-key}, \textit{test-key'}$) into the next available positions in $curr$.

\noindent (5)~\textbf{Advance.} Repeat Steps 2--4 until either (i) the node is full, or (ii) $sublist_i$ is exhausted. If keys remain and $k' >$ node max of $curr$, advance $curr \leftarrow curr.\texttt{next}$ and repeat from Step 1. If (i), and $sublist_i$ is not exhausted -- split the node and continue from Step~1 at the appropriate node.

}

\textbf{Hybrid ST and TL Algorithms} (\texttt{ST-Hybrid} \& \texttt{TL-Hybrid}).
Hybrid insertions combine shift-right and bulk strategies depending on the node fill state and workload.

\subsection{Deletion Kernels}
\label{sec:deletionalgs}

We evaluate three kernels: (i)~\texttt{ST-Shift-Left}, (ii)~\texttt{TL-Shift-Left}, and (iii)~\texttt{TL-Bulk} deletions. We do not employ a tombstone strategy because it associates a performance penalty with another operation (insertions or queries) and incurs memory overhead. ST-\texttt{Shift-Left} and \texttt{TL-Shift-Left} deletions mirror the \texttt{Shift-Right} insertion strategy. \texttt{TL-Bulk} is outlined below and illustrated with an example in Table~\ref{tab:TL_bulk_delete}. The steps are:

{\small
\noindent (1)~Each tile completes \texttt{Common Steps} from Section \ref{sec:Alg-commonsteps} and proceeds with a $sublist_i$ for bucket $b_i$.

\noindent (2)~Each thread in $tile_i$ loads one key from $curr$.

\noindent (3)~For each delete key $dk$, threads compare their stored key with $dk$, marking matches; a counter tracks the \textit{del count}.

\noindent (4)~A compaction step shifts surviving keys left; the shift distance per thread depends on the number of preceding deletions.

\noindent (5)~All \textit{del count} keys are removed, the node size is reduced. Empty nodes are removed from the chain and made available for subsequent insertions.

\noindent (6)~Tile continues with the remaining delete keys from $sublist_i$, traversing node-link pointers in bucket $b_i$ as needed.

}

\subsection{Fully Concurrent Kernel over Mixed Batches}
\label{sec:flix-fully-concurrent}

We combine the above ST and TL kernels to offer fully concurrent execution over \textit{mixed batches} of point queries, inserts and deletes, using shift-right for insertions and shift-left for deletions.
Batches, which contain key and operation-type pairs, are sorted by key, so threads or tiles can binary search for relevant keys.
A tile then makes three passes over its bucket, inserting, deleting, then querying.
Operations in a bucket can progress independently of other buckets, so at any moment one warp may be inserting into its bucket while another performs deletions and a third answers queries.
Because of our bucket-oriented computing, neither locks nor atomics are needed, except to allocate and free nodes.

\section{Update Experiments }
\label{sec:update-experiments}

\begin{figure*}[t]
    \centering
    \begin{subfigure}[b]{0.24\textwidth}
        \centering
        \includegraphics[width=\linewidth]{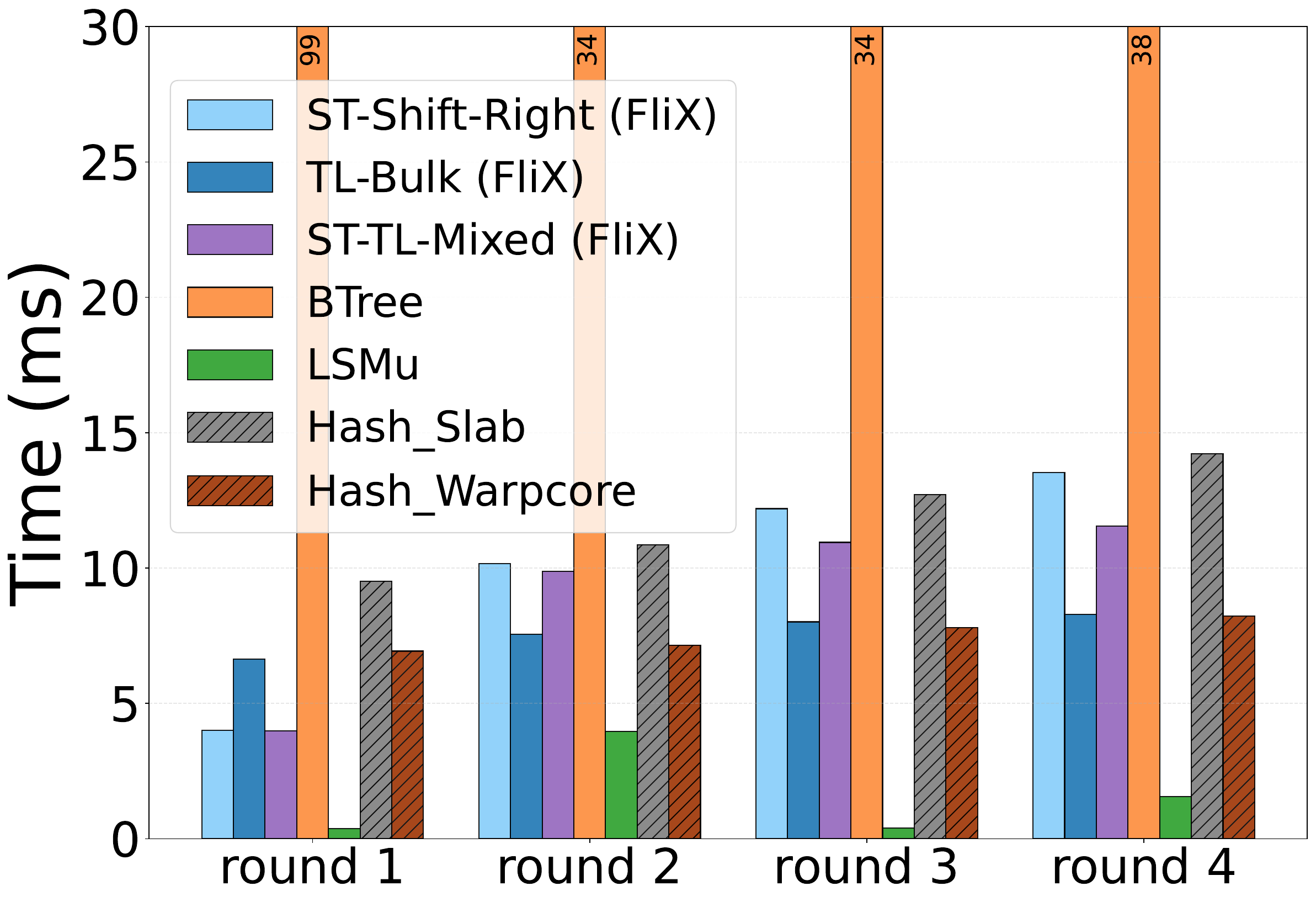}
        \caption{Insert Time: $X=90$, $Y=90$}
    \end{subfigure}
    \hfill
    \begin{subfigure}[b]{0.24\textwidth}
        \centering
        \includegraphics[width=\linewidth]{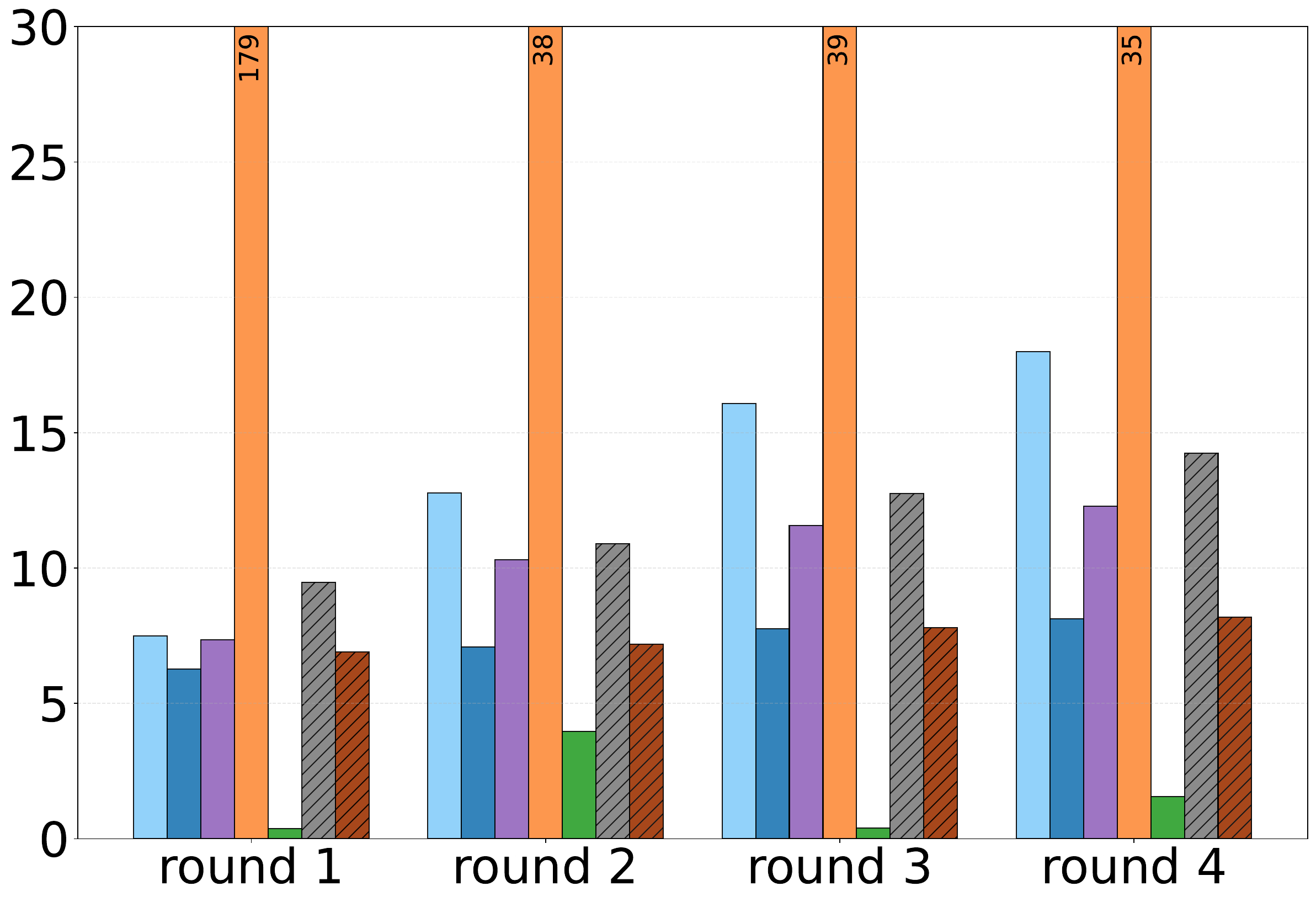}
        \caption{Insert Time: $X=25$, $Y=90$}
    \end{subfigure}
    \hfill
    \begin{subfigure}[b]{0.24\textwidth}
        \centering
        \includegraphics[width=\linewidth]{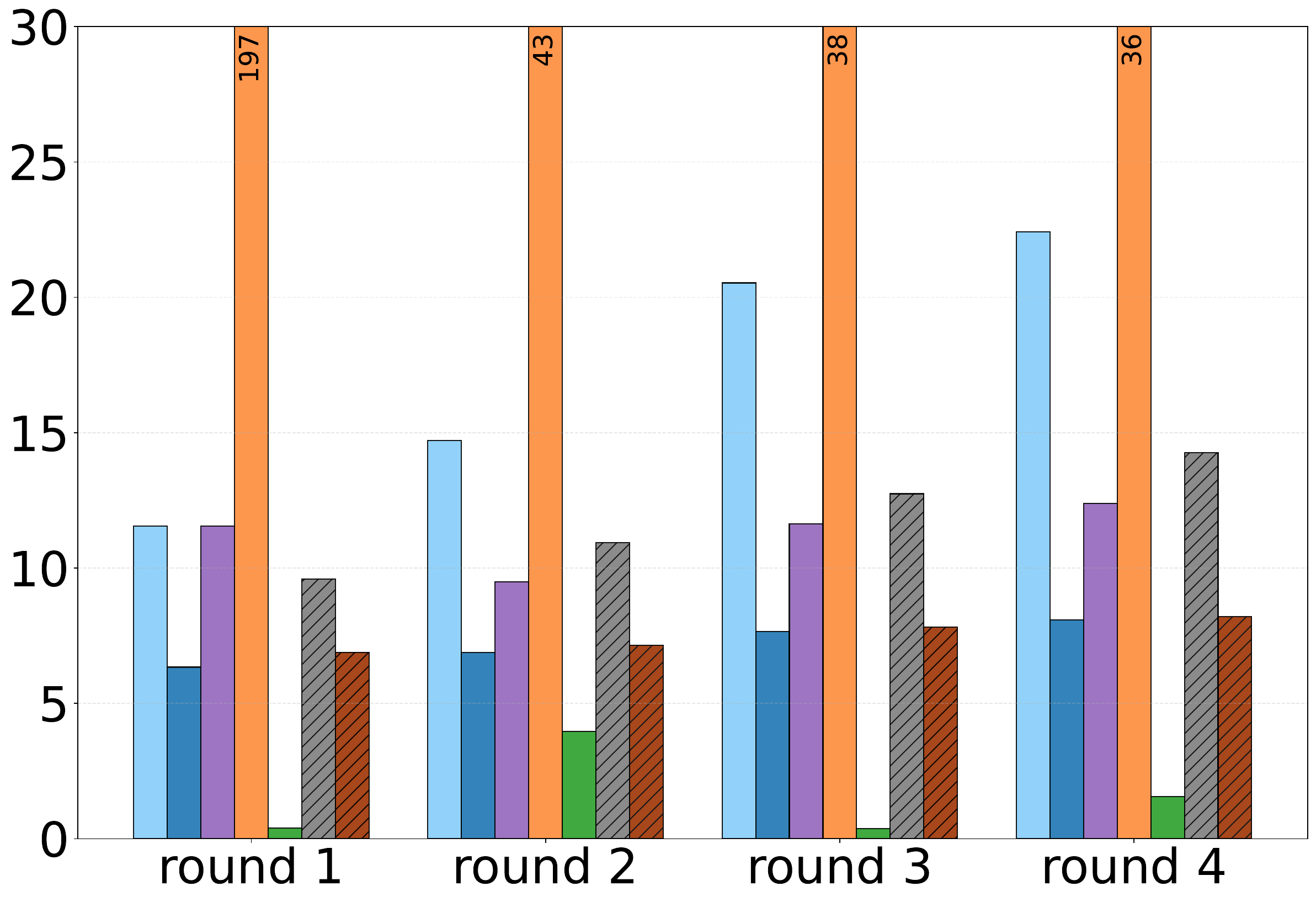}
        \caption{Insert Time: $X=6$, $Y=90$}
    \end{subfigure}
    \hfill
    \begin{subfigure}[b]{0.24\textwidth}
        \centering
        \includegraphics[width=\linewidth]{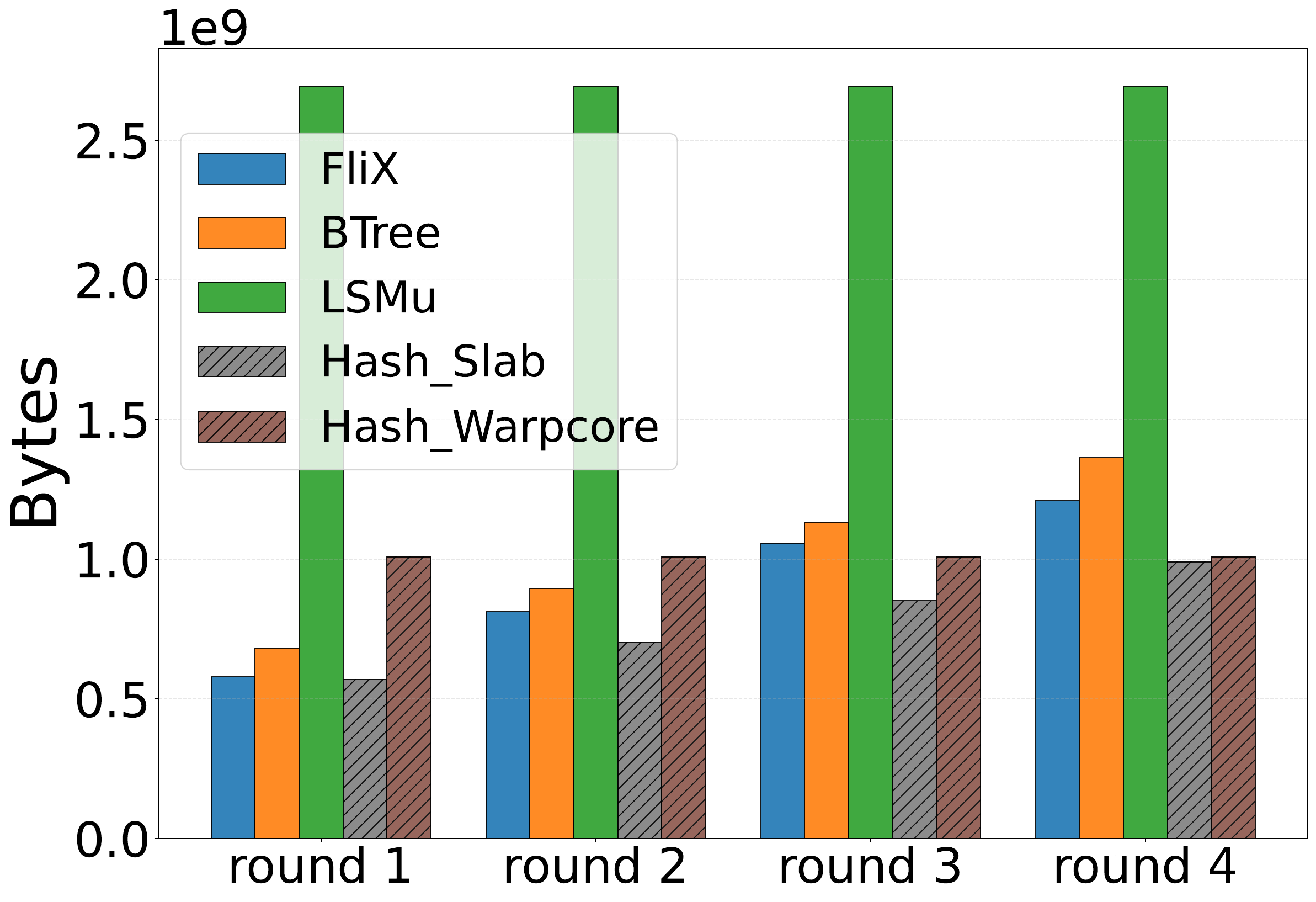}
        \caption{Memory Footprint/Round}
    \end{subfigure}

    \caption{ \small
        Experiment with 4 consecutive rounds of insertions across 3 workloads of varying key densities. The \texttt{TL-Bulk} implementation consistently outperforms the~\btree{} and \slabhash{} and is comparable to~\warphash{}.~\lsmu{} has the strongest insert performance in all experiments
        at the cost of significantly higher memory consumption, illustrated in (d).
    }
    \label{fig:baselines}

    \Description{varying workloads: FliX insertions compared to baselines }
\end{figure*}

\begin{figure}[tb]
    \centering

    \setlength{\abovecaptionskip}{2pt}
    \setlength{\belowcaptionskip}{0pt}

    \begin{subfigure}[b]{0.48\columnwidth}
        \centering
        \includegraphics[width=\linewidth]{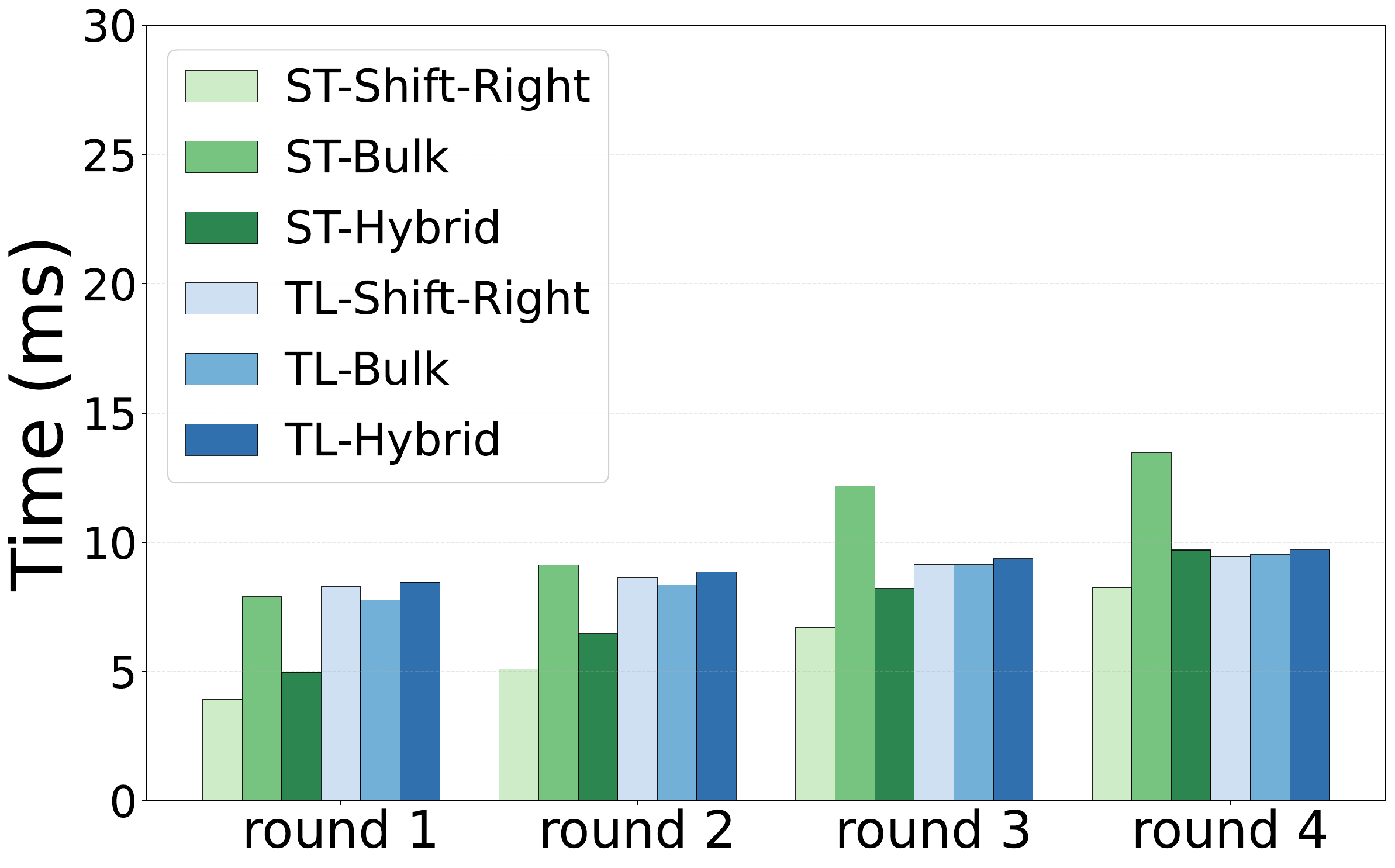}
        \caption{Uniform Insertions}
    \end{subfigure}
    \hfill
    \begin{subfigure}[b]{0.48\columnwidth}
        \centering
        \includegraphics[width=\linewidth]{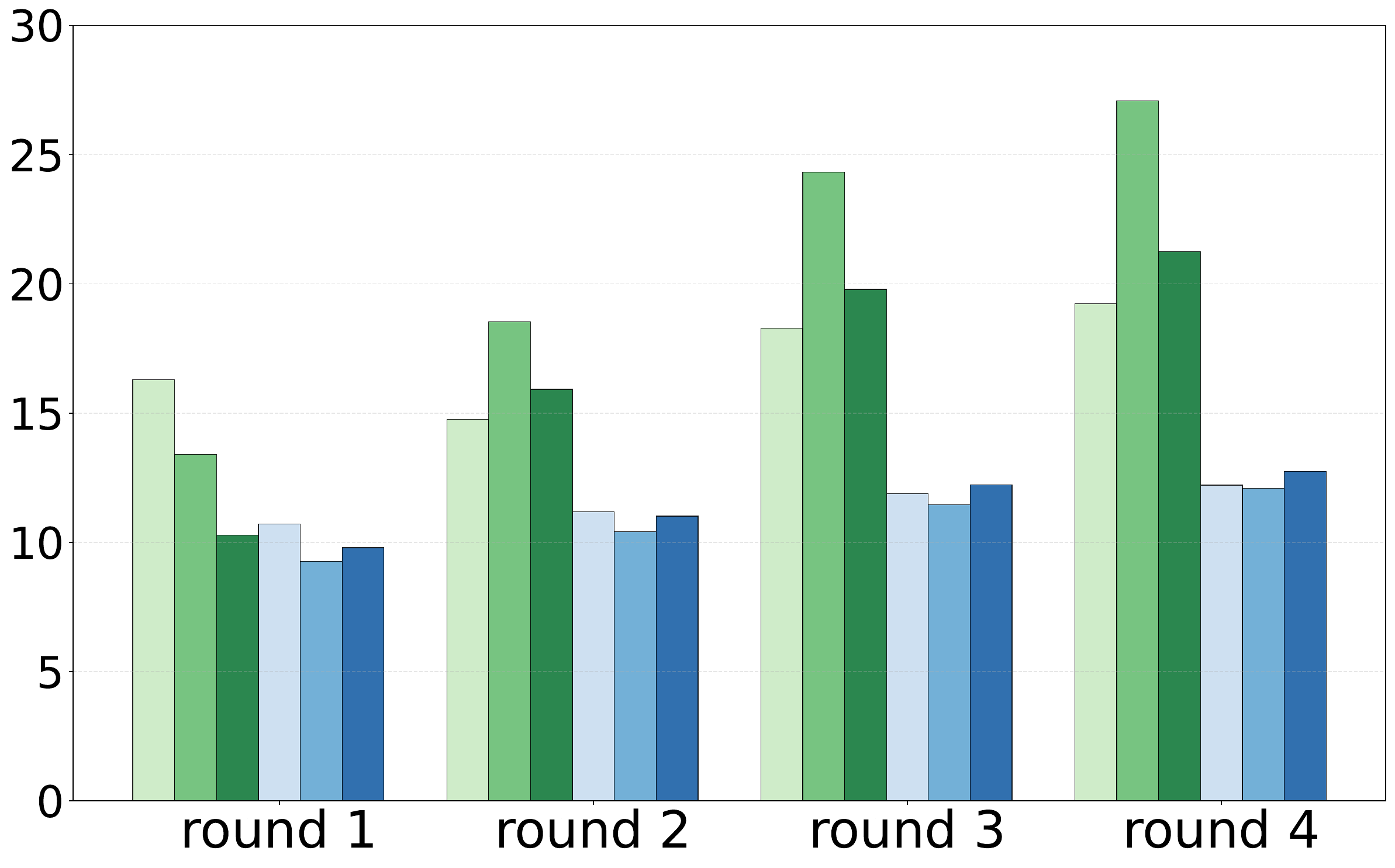}
        \caption{Dense Insertions}
    \end{subfigure}

    \caption{ \small Comparison of \flix{} insert kernels (ST or TL). (a) Uniform distribution with 100\% growth, (b) Dense distribution with 200\% growth.}
    \vspace{-2mm}
    \label{fig:use-case-single-threaded-inserts}
    \Description{lighter insertion workload FliX}

\end{figure}

\subsection{Baselines}
\label{sec:baselines-for-experiments}

In this study, we compare \flix{} against 4 GPU CDS baselines: the ~\btree{}~\cite{lit:btree1}, \slabhash{}~\cite{lit:hash-slabhash}, \warphash{}~\cite{lit:warpcore}, and \lsmu{}, our extended variant of the GPU \lsmtree{} by Ashkiani et al.~\cite{lit:lsmtree}.
For the \btree{} we use the recommended node size of 15 keys%
~\cite{lit:btree1}.
We initialize both hash tables at 80\% load factor to balance memory usage and performance.
The two hash tables serve as a reference for the performance of unordered structures, and are not direct competitors for \flix{}.
Rather, they allow us to quantify the price an ordered structure pays for maintaining order.
\lsmu{} is our improved version of the GPU \lsmtree{}~\cite{lit:lsmtree}, differing in three ways.
First, a deletion overwrites the existing key's value with a tombstone in place rather than inserting a new entry, so each key appears at most once and binary search within a level still applies.
Second, range queries are processed with one cooperative tile per query.
Third, we add \textit{next-larger} (successor) queries.
Insertions are unchanged, and each modification either removes an overhead or adds an operation, so \lsmu{} is at least as strong a baseline as the original on every workload we measure.

\subsection{Experimental Setup}
Experiments run on an NVIDIA A6000 GPU (84 SMs) with an Intel Xeon Gold 5220 host (72 physical cores).
The build phase of~\flix{} is described in Section~\ref{sec:flixBuild}.
Unless stated otherwise, each data structure is built from $2^{25}$ ($\approx$33 million) uniformly distributed \textit{key-rowID} pairs.
Update experiments then apply 4 rounds of batched insertions of $16.8$ million pairs each, tripling the size of the data structure, and rounds of batched deletions that remove them.

Update workloads concentrate updates in part of the key range, controlled by two parameters.
The \textit{\textbf{key range percentage}}~$X$ is the fraction of the key range that receives most updates, and the \textit{\textbf{update percentage}}~$Y$ is the fraction of update keys it receives.
For example, with $X=12\%$ and $Y=90\%$, $12\%$ of the key range receives $90\%$ of the update keys, so a set of \textit{hot} buckets receives most updates in every round; the remaining $10\%$ are spread uniformly over the rest of the key range to avoid caching bias.
We vary $X$ from $90\%$ (with $Y=90\%$, a uniform distribution) down to $3\%$.
Section~\ref{sec:highlyskewed-queries} does the same for
queries, with $Y=100\%$ and $X$ down to $0.0001\%$.

Reported times measure GPU execution only.
\textit{Costs that are the same for all data structures are excluded}.
These include CPU--GPU transfers, kernel launch overhead, and result validation on the host.
Batches are sorted by key on the GPU before each experiment, and reported times exclude this sort for all data structures; sorting a batch of $2^{24}$ keys takes $\approx 1.7$ ms (Table~\ref{tab:avg_sort_time_a6000}).
The exception is Section~\ref{sec:unsorted}, where all structures receive unsorted batches and \flix{} is charged for sorting them.

\begin{figure}[tb]
    \centering
    \begin{subfigure}[b]{0.48\columnwidth}
        \centering
        \includegraphics[width=\linewidth]{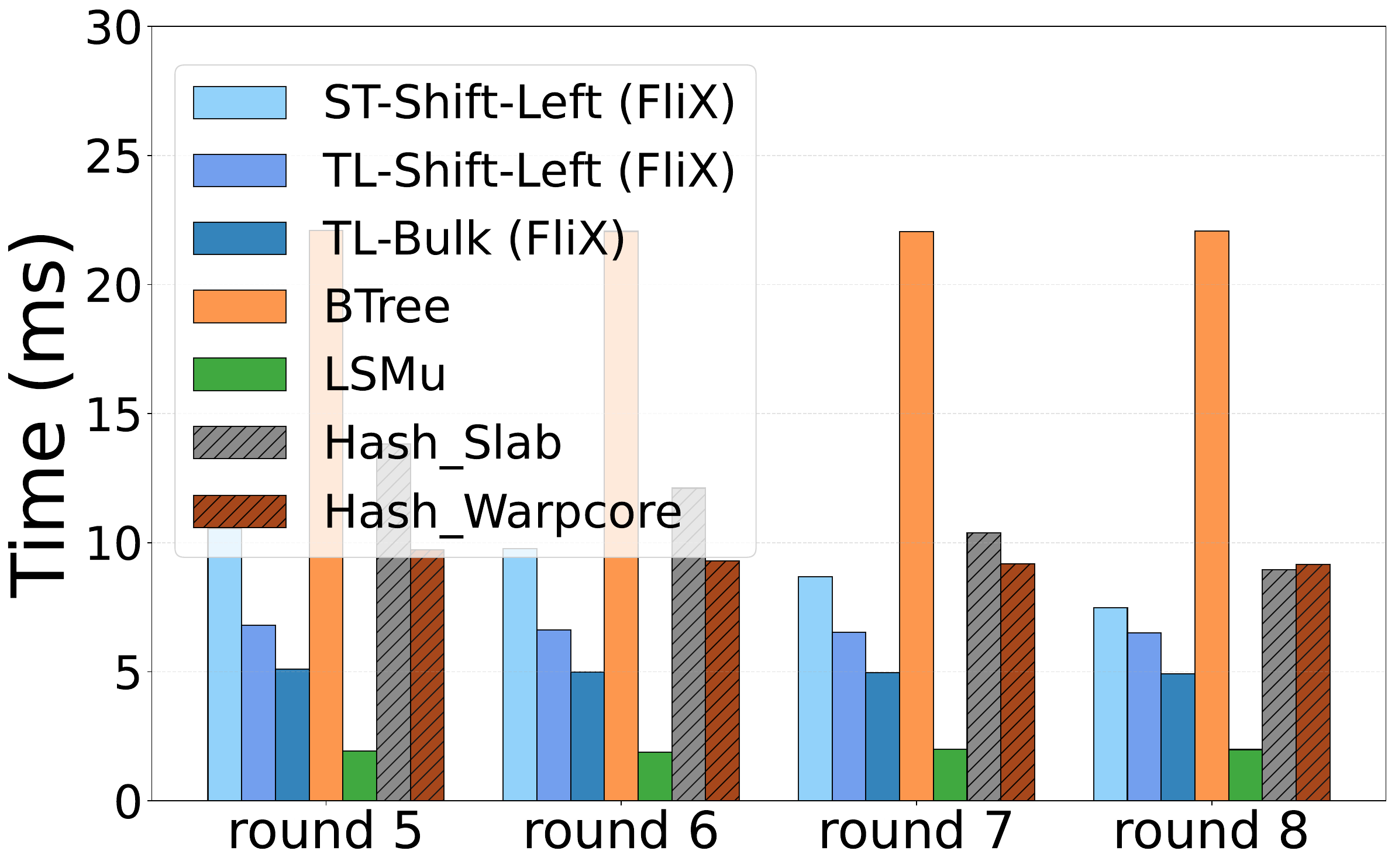}
        \caption{Deletions: $X=90,\;Y=90$}
    \end{subfigure}%
    \hfill
    \begin{subfigure}[b]{0.48\columnwidth}
        \centering
        \includegraphics[width=\linewidth]{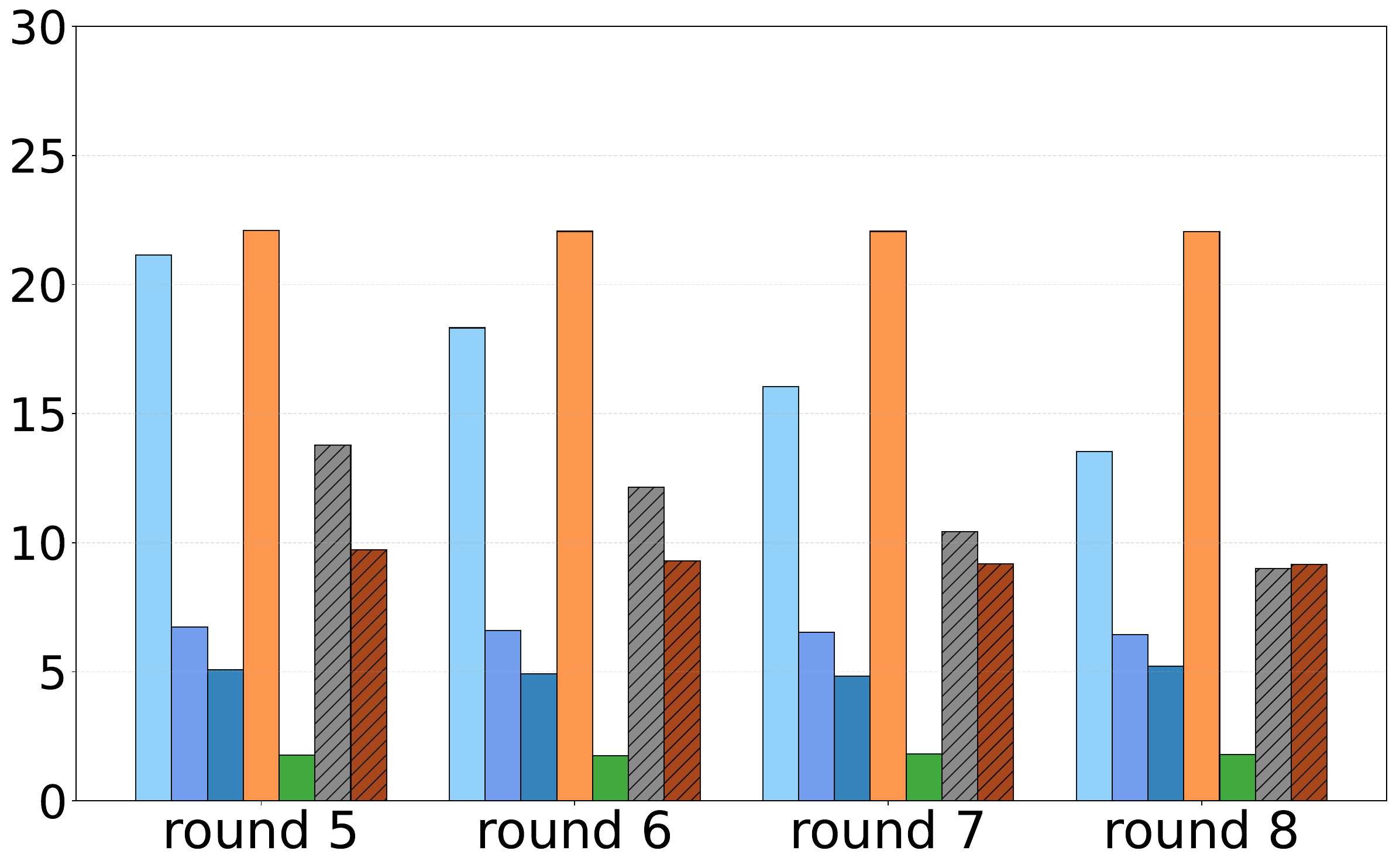}
        \caption{Deletions: $X=6,\;Y=90$}
    \end{subfigure}

    \caption{ \small Comparing~\flix{} deletion kernels against baselines.}
    \label{fig:baselines-del-exp}
    \Description{Deletion performance in FliX}

\end{figure}

\begin{figure*}[tb]
    \centering
    \setlength{\tabcolsep}{4pt}
    \begin{subfigure}[b]{0.48\textwidth}
        \centering
        \includegraphics[width=\linewidth]{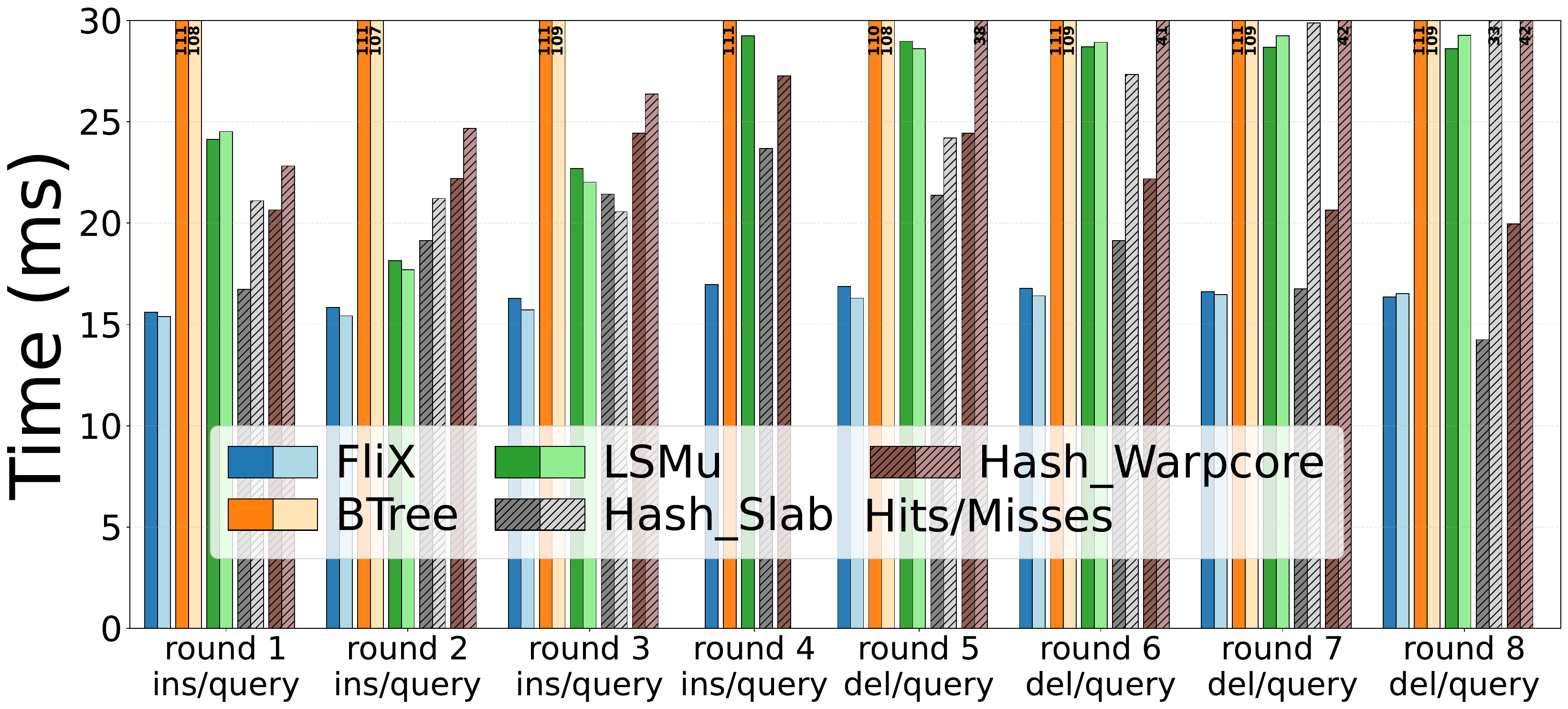}
        \caption{Query Time (Hit/Miss)}
    \end{subfigure}
    \hfill
    \begin{subfigure}[b]{0.48\textwidth}
        \centering
        \includegraphics[width=\linewidth]{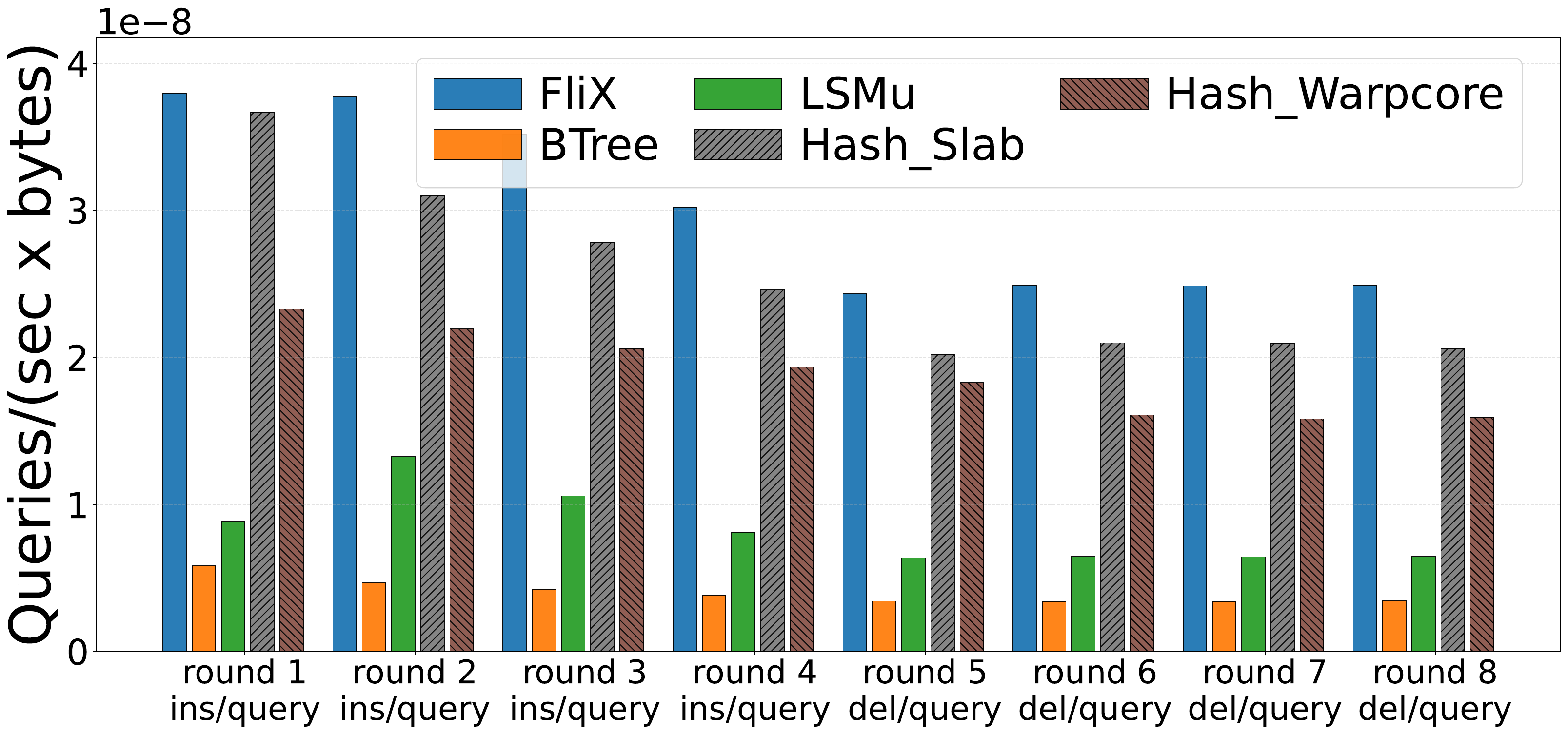}
        \caption{Query Throughput / Memory Footprint}
    \end{subfigure}

    \caption{  \small \flix{} (TL) query performance against baselines following rounds of insertions and deletions ($X=90$, $Y=90$). Figure (a) shows query performance (latency in ms) after each update. Figure (b) reports query throughput normalized by memory usage.}
    \Description{Query performance after consecutive updates.}
    \label{fig:probe-hit-miss-tpmf}

\end{figure*}

\subsection{Formative Experiments}
\label{sec:Formative experiments}
Our formative experiments evaluate a cross product of configuration settings of \flix{} for each update kernel to identify the most performant combinations.
Hybrid algorithms that switch between \textit{bulk} and \textit{shift-right} insertions (\texttt{ST-Hybrid} \& \texttt{TL-Hybrid}) were not performant overall,
so we exclude them from further consideration.

\textbf{Node Size and Threads per Block}.
For each insertion kernel, we vary the $nodeSize$ over 8, 14, 16, and 32 keys (a 14-key node, with its metadata, fills one cache line) and the $threadsPerBlock$ over 1024 (A), 512 (B), 256 (C), and 128 (D), giving 16 variants per kernel, and evaluate each variant under several workloads.
Since each thread of a TL kernel handles one slot of a node, the tile size is the smallest power of two greater than or equal to $nodeSize$.
Figure~\ref{fig:full-heatmaps} summarizes the results as a heat map, in which each \textit{\textbf{cell}} is one kernel variant and darker blue means lower insertion time.

\textbf{Best Overall Configuration}.
Each \textbf{\textit{row}} of Figure~\ref{fig:full-heatmaps} is one round of insertions under one workload, and each cell is normalized to the best variant in its row, so a score of 2 means twice the time of the best (capped at 2).
\texttt{TL-Bulk} with $nodeSize=2^5$ and $threadsPerBlock=128$ scores best overall, with the most 1.0 scores.
Tiled kernels generally favour large nodes and few threads per block.
A 32-key node matches a full warp, avoiding the scheduling overhead of smaller tiles, and for \texttt{TL-Bulk} it also leaves more free space per node, so more keys are inserted in each bulk operation.

\textbf{When Single-Threaded Kernels Win}.
The heat map also shows that ST kernels, particularly \texttt{ST-Shift-Right} with small nodes, remain competitive when there is little work per bucket, since a TL kernel then leaves most threads of each tile idle and still pays for synchronization.
Figure~\ref{fig:use-case-single-threaded-inserts}~(a) isolates this case: with a build size of $2^{25}$ ($\approx$~4.2 million buckets) and $8.38$ million uniform insertions per round (100\% growth over 4 rounds), each bucket receives only 2--3 keys per round.
In the denser workload of Figure~\ref{fig:use-case-single-threaded-inserts}~(b), where $25\%$ of buckets receive most insertions, \texttt{TL-Bulk} is faster.
Overall, \texttt{ST-Shift-Right} performs best with small nodes ($2^3$ keys), near-uniform distributions, and at most 2--5 insertions per bucket per round.

\textbf{Mixed Insertions}.
\texttt{ST-Shift-Right} is faster in the first round under near-uniform workloads and \texttt{TL-Bulk} in later rounds, so our \texttt{ST-TL-Mixed} kernel switches from the former to the latter after the first round.
Unlike the hybrid kernels, it combines kernels \textit{across} the ST and TL categories rather than switching strategies within one.
It uses $nodeSize=2^3$ and $threadsPerBlock=256$, fixed at build time (results in Section~\ref{sec:comparebaselines}). Mixed insertions should not be confused with \textit{mixed batches} in the concurrent setting which is discussed in Section~\ref{sec:concurrent-updates-queries}.

\subsection{Comparison with Baselines}
\label{sec:comparebaselines}
\textbf{Insertions}.
Figure~\ref{fig:baselines} compares the \flix{} insertion kernels against the baselines.
\texttt{TL-Bulk} consistently outperforms the GPU~\btree{} and~\slabhash{}, by $4.3$--$6.3\times$ and $1.6\times$ on average after the first round, respectively, and is comparable to~\warphash{}. Including the first round, \texttt{TL-Bulk} outperforms the~\btree{} by $8\times$.

\texttt{ST-Shift-Right} and \texttt{ST-TL-Mixed} degrade as the data structure grows, especially under the denser workloads of Figure~\ref{fig:baselines}~(b)--(c).
\texttt{ST-TL-Mixed} suffers because node size is fixed at build time, so after switching, \texttt{TL-Bulk} runs with the small nodes chosen for \texttt{ST-Shift-Right} rather than its own best configuration.
\lsmu{} has the fastest insertions of all structures, leveraging fast GPU sort-and-merge, but its merges need large auxiliary buffers: Figure~\ref{fig:baselines}~(d) shows that \flix{} uses $2.8\times$ less memory than \lsmu{}, and consistently less than the \btree{}.

\textbf{Deletions}.
A heat map analogous to Figure~\ref{fig:full-heatmaps} (not shown) identifies \texttt{TL-Bulk} as the best deletion kernel across all workloads, with the same parameters as for insertions ($nodeSize=2^5$, $threadPerBlock=128$).
Figure~\ref{fig:baselines-del-exp} shows the 4 rounds of deletions that follow the insertions of Figure~\ref{fig:baselines}.
\texttt{TL-Bulk} outperforms the other \flix{} kernels and every baseline except~\lsmu{}, deleting on average $4.4\times$ faster than the~\btree{}, $2.2\times$ faster than~\slabhash{}, and $1.8\times$ faster than~\warphash{}.
Unlike the hash tables and~\lsmu{}, which leave tombstones, \flix{} physically removes deleted keys and reclaims empty nodes immediately.
Underfull nodes are merged later by restructuring.

\section{Query Experiments}
\label{sec:query-experiments}

Next, we analyze the query performance of~\flix{} while keeping insertion and deletion rounds part of the experiments. As the insertion and deletion kernel, we use \texttt{TL-Bulk} in the same configuration as evaluated before. For point queries, we also support both an ST and a TL variant of the principle described in Section~\ref{sec:queries}.
Based on the formative experiments of the previous section, we use $nodeSize=2^3$ for~\flixst{} and $nodeSize=2^5$ for the tiled~\flix{}, each with its optimal settings.
However, as we found the ST variant to be performant only when the batch size is small and the number of queries per bucket is low, we primarily focus on the TL implementation for point queries, which we simply denote as~\flix{} in the following plots. The ST variant, which we include for completeness wherever meaningful, is denoted~\flixst{}.

In subsequent experiments, each round consists of a batch of insertions or deletions, followed by two batches of $2^{28}$ point queries, unless stated otherwise. The first query batch contains only keys that are hit, while the other contains only keys that are misses. If all keys of the experiment have been inserted already and no queries that miss can be generated anymore, we exclude the miss bar in the corresponding round.

\begin{figure}[tb]
    \centering
    \includegraphics[width=\columnwidth]{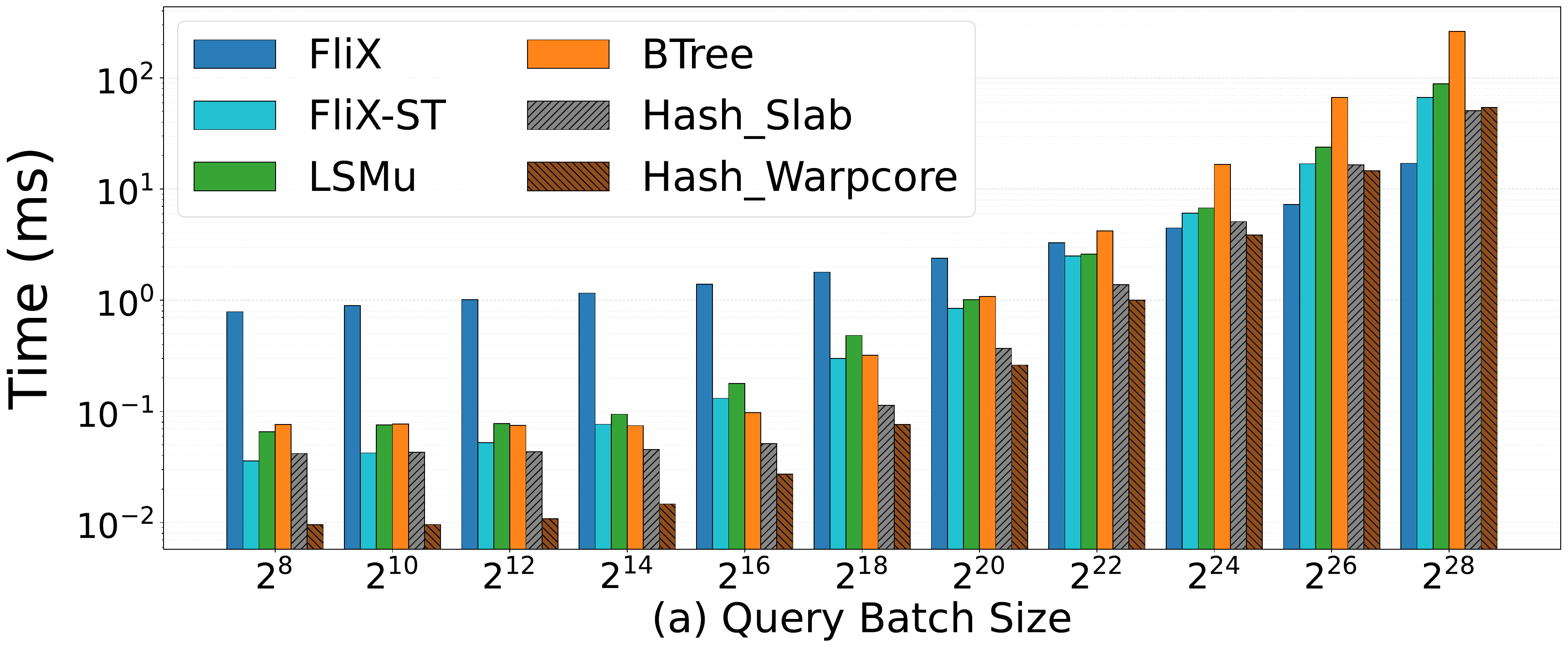}
    \caption{\small Query latency for varying query batch sizes, with a fixed build size of $2^{25}$.}
    \label{fig:vary-batch-size}
    \Description{varying GPU batch and build sizes.}
    \vspace{-2mm}
\end{figure}

\subsection{Query Latency vs Memory Footprint}
\label{sec:query-qtmf-results}

We first inspect the query latency in the presence of updates for all data structures, where we assume that queries are already sorted. Additionally to raw latency, we are interested in the query throughput per memory footprint (QTMF), which shows how each method ``buys'' its performance by consuming memory, i.e., by building auxiliary internal structures to speed up lookups. Figure~\ref{fig:probe-hit-miss-tpmf} shows query latency and throughput per memory footprint for an initial build size of $100$~million keys with an overall growth factor of $100\%$, achieved through four successive rounds of uniformly distributed insertions. Subsequently, the same keys are removed in four rounds of deletions shrinking the structure to its original size. For this experiment, we test uniform modifications to the data structure ($X=90$ and $Y=90$); we will test skewed distributions in Section~\ref{sec:highlyskewed-queries}. In Figure~\ref{fig:probe-hit-miss-tpmf}~(a), we observe that in terms of query latency, \flix{} clearly outperforms both ordered data structures, on average by~$6.5\times$ over the~\btree{} and by~$1.5\times$ over~\lsmu{}. Even the hash tables are outperformed in a majority of cases, especially under misses, where \flix{} outperforms both hash tables on all-miss queries by $2$--$2.6\times$.

Figure \ref{fig:probe-hit-miss-tpmf} (b) shows the QTMF results, where higher values indicate better throughput per memory footprint. Here, we observe that the \btree{} and \lsmu{} do not perform well under this metric due their high memory consumption. \flix{} on the other hand remains strong even against unordered baselines.

\subsection{Varying Batch Size}
\label{sec:vary-batch-sizes}

So far, we fired a whole batch of $2^{28}$ queries. In the following, we will vary the batch size from $2^{8}$ to $2^{28}$ to study the effect on all methods. Here, we include \flixst{} to see whether it handles potential under-utilization better than the tiled variant. Figure~\ref{fig:vary-batch-size} shows the results averaged over ten rounds of updates, with a growth rate of 200\%, using a \textit{$X=25$} and \textit{$Y=90$} workload. In Figure~\ref{fig:vary-batch-size}~we vary the query batch size (100\% hits) while keeping the build size fixed at $2^{25}$~keys. The results reveal that the ideal choice between ST and TL depends on batch size, where the breakpoint between the two is at about $2^{20}$; at this batch size, there are roughly 4 queries per bucket for~\flixst{} and 16 queries per bucket for~\flix{} (query batch size / number of buckets). With smaller batch sizes, e.g., $2^8$, buckets receive less than one query on average, so a tile sits mostly idle.

\begin{figure}[tb]
    \centering
    \includegraphics[width=\columnwidth]{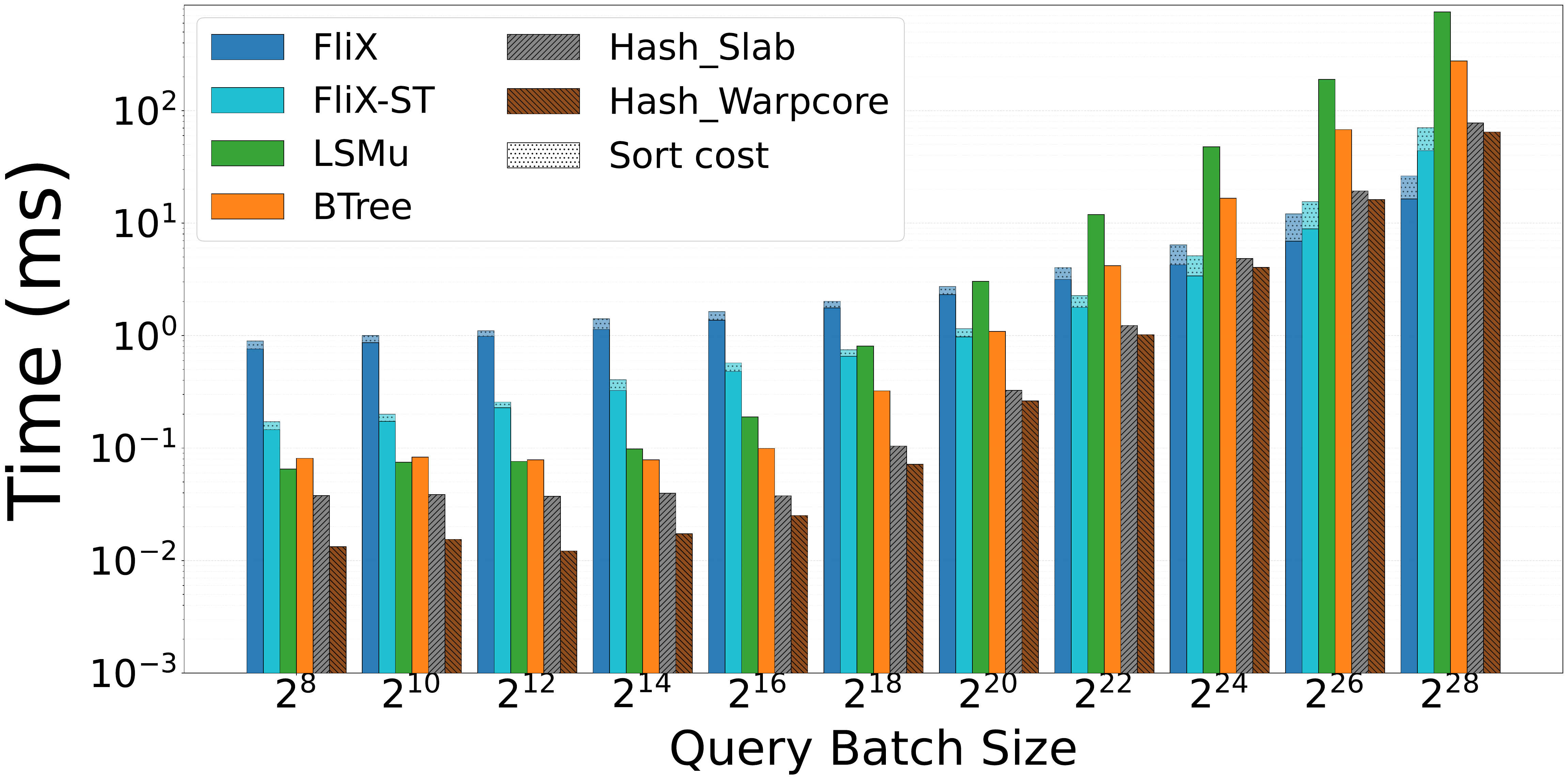}
    \caption{\small Unsorted query performance (log scale), baselines vs~\flix{}; build size $2^{25}$,
    cost of sorting shown as a stacked bar for~\flix{}.}
    \label{fig:unsorted}
    \Description{Unsorted performance: FliX against baselines}
\end{figure}

\subsection{Unsorted Queries}
\label{sec:unsorted}

Next, we inspect the case when queries arrive unsorted. For \flix{} this means that a sorting step must be performed before the actual query answering starts, which we will report in the measurements, whereas the baselines directly fire the batch of queries as received.

The experiment uses a build size of $2^{25}$ keys and again varies the query batch size, comparing both the default tiled~\flix{} and~\flixst{} implementations against the baselines. Figure~\ref{fig:unsorted} illustrates the average query performance across 6 rounds of unsorted \textit{all-hit} queries. The results indicate that even when the cost of sorting is included, the tiled~\flix{} implementation outperforms all baselines at large batch sizes. At a batch size of $2^{28}$, it is $17.6\times$ faster than~\lsmu{}, $6.5\times$ faster than the~\btree{}, and $1.8\times$ and $1.5\times$ faster than~\slabhash{} and~\warphash{}, respectively.

\begin{figure}[tb]
  \centering
  \begin{subfigure}{0.48\columnwidth}
    \centering
    \includegraphics[width=\linewidth]{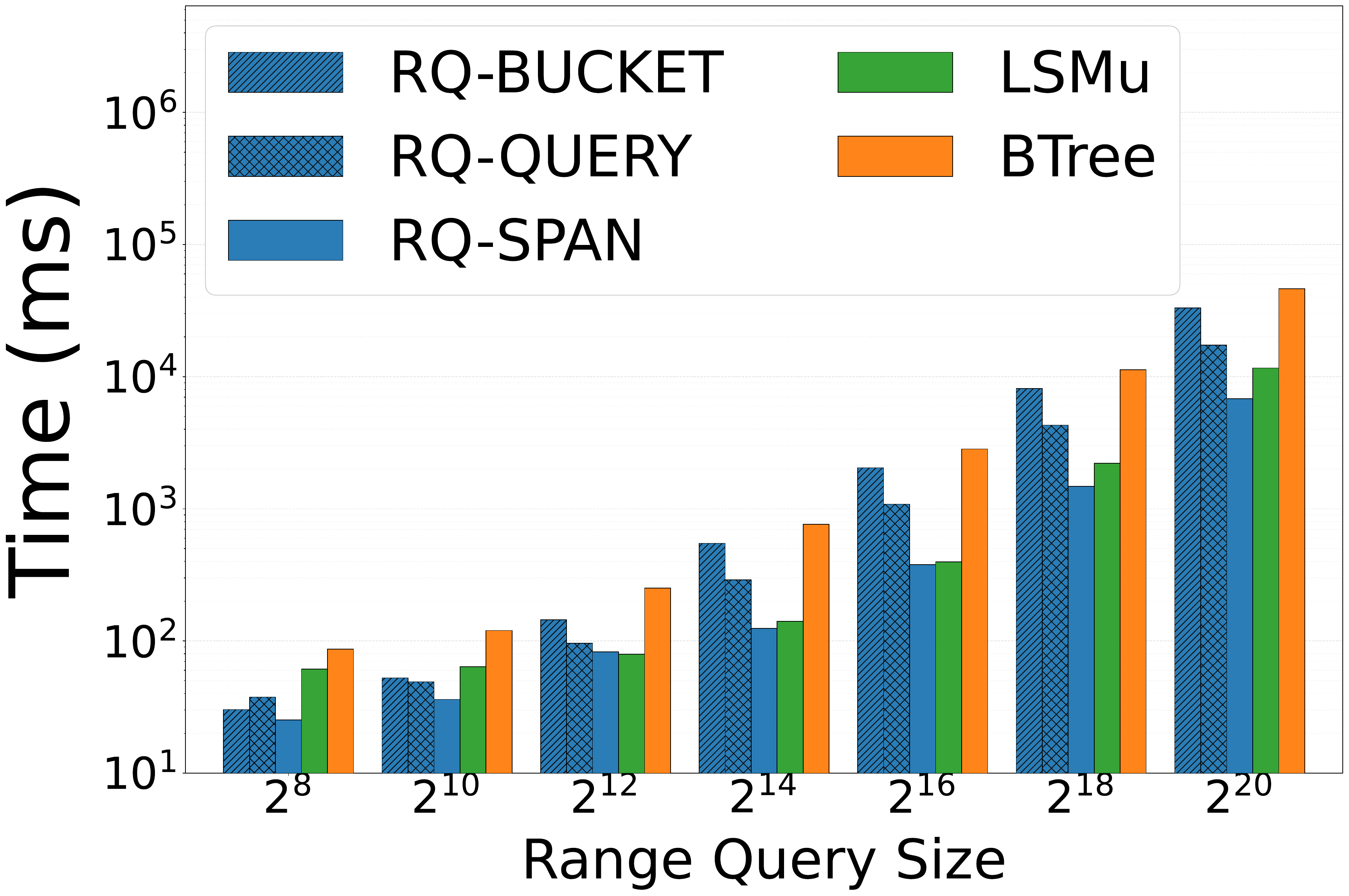}
    \caption{Range scan (log scale)}
  \end{subfigure}%
  \hfill
  \begin{subfigure}{0.48\columnwidth}
    \centering
    \includegraphics[width=\linewidth]{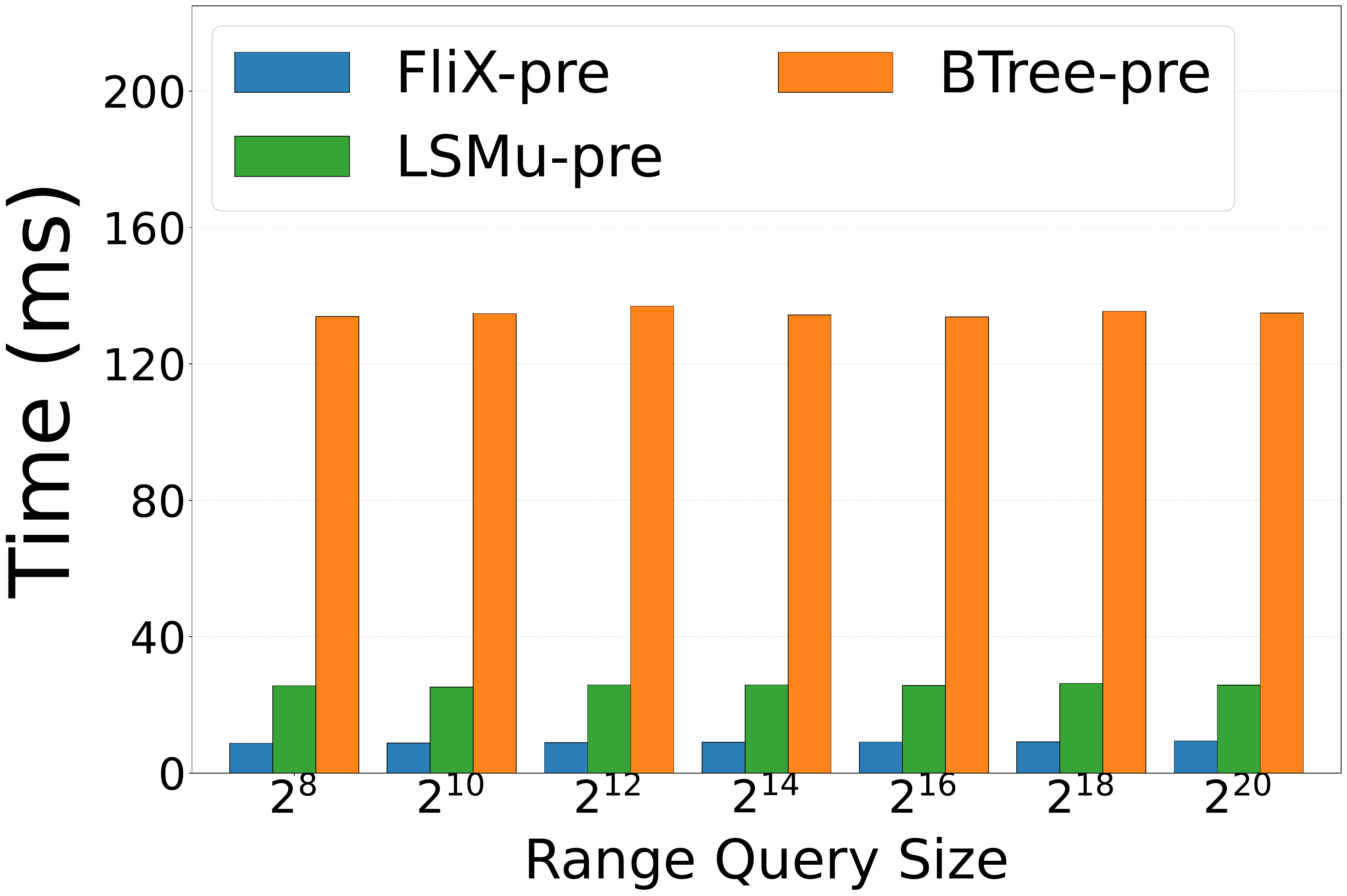}
    \caption{Precompute sums (linear)}
  \end{subfigure}

  \caption{\small Range Query results where 67 million uniformly generated range queries ($2^{26}$) are performed per round following insertions.}

  \label{fig:rq_vary_size}
  \Description{Range queries in FliX}
\end{figure}

\subsection{Range Queries}
\label{sec:range_queries}

Let us next evaluate the range query performance of \flix{} in Figure~\ref{fig:rq_vary_size}. To do so, we vary the size of the fired range queries from $2^8$ to $2^{20}$ and compare our three range query variants for \flix{}, namely \rqbucket{}, \rqquery{}, and \rqspan{}, against \btree{} and \lsmu{}.

Figure~\ref{fig:rq_vary_size} (a) shows the results for range scans on a logarithmic scale, in which each range query computes a sum over the retrieved rowIDs of the range. From the results, we can see that out of our variants, \rqspan{} performs best across all configurations. Further, \rqspan{} is also able to outperform the baselines for almost all range sizes, due to its flat consecutive structure that fits the nature of range queries. In contrast, the \btree{} has to traverse its index layer and~\lsmu{} must go through all of its levels.

Additionally, in Figure~\ref{fig:rq_vary_size} (b), we test how the individual structures could evaluate range queries over precomputed aggregates, as described in Section~\ref{sec:range-queries}. To have a fair comparison, we implemented a corresponding support not only for \flix{}, but also for \btree{} and \lsmu{}. The results show that \flix{} manages to clearly outperform its competitors in this regard.

\subsection{Skewed Queries}
\label{sec:highlyskewed-queries}
Recall, a distributional shift can occur when the data structure is built on one distribution, and subsequent operations adhere to a very different distributional spread. This may lead to many tiles remaining inactive, while others receive a large number of operations.
We assess the impact of engaging \textit{one tile per bucket}, under skewed workload distributions in Figure~\ref{fig:skewedworkloads-flix}.
The experiment uses a build size of $2^{25}$, resulting in $\approx$ 2 million buckets, where $100\%$ of query keys are drawn from the given \% of the distribution. The query size per round is $2^{26}$ keys, and the average query response time is reported across 8 rounds of query-only workloads. The number of buckets engaged per experiment is labelled over top: at $0.0001$ of the distribution, only 3 buckets are receiving queries. This is constant across varying query batch sizes; the difference being an increase in duplicate keys as the query batch size increases. We observe that only a small degradation in query performance occurs even when the query workload is concentrated to $1\%$ or $0.1\%$ of the distribution. A more severe penalty occurs when the query workload is concentrated to a fractional proportion of the data structure, since GPU warp occupancy is then concentrated to only a few SMs. For practical purposes, this appears to be a performance penalty we can live with; our results indicate GPU tile utilization is highly robust to distributional shifts in the workload.

\begin{figure}[tb]
 \includegraphics[width =\columnwidth]{Figures/average_probe_time_log_scale_with_total_buckets_used_final_v3.pdf}

  \caption{ \small~\flix{} experiments with a build size of $2^{25}$; 100\% of query keys are drawn from increasingly dense regions of the distribution.
  Query time degrades severely when the proportion of buckets is very small ($0.01\% - 0.0001\%$). }%

  \label{fig:skewedworkloads-flix}
  \Description{skew performance in queries: FliX}

\end{figure}

\begin{figure}[tb]
  \centering
  \includegraphics[width=\columnwidth]{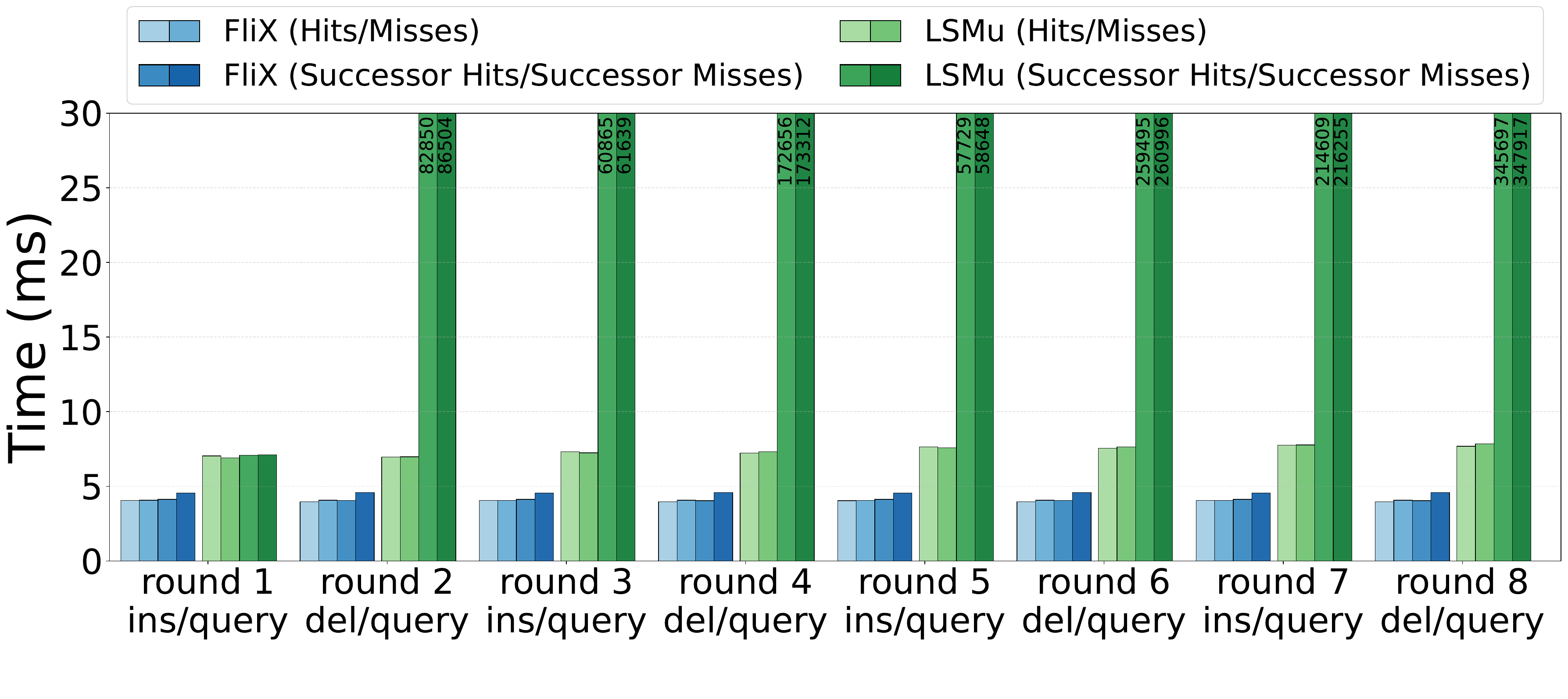}

  \caption{\small Successor results comparing the \lsmu{} and \flix{}.}
  \label{fig:successor}
  \Description{Successor query results comparing \lsmu and \flix{}.}

\end{figure}

\subsection{Successor Queries}
\label{sec:query-experiments-successor}

A successor query returns the smallest key in the data structure that is greater than or equal to a search key. For a search key $k$, the successor operation returns the smallest key $k'$ satisfying $k' \ge k$.
Figure~\ref{fig:successor} reports successor-query performance for~\lsmu{} and~\flix{}. As the deletion rate increases,~\lsmu{} performance degrades markedly. By round~8,~\flix{} successor query time is $\approx$ 69000$\times$ lower than~\lsmu{}. This heavy degradation in~\lsmu{} performance stems from tombstone handling: successor queries must traverse levels in the tree while skipping tombstones, which effectively can become a linear scan\textit{ by each thread in each level} to locate the next active key.
Tombstones impose an unavoidable cost in GPU CDSs;
they increase work in queries and insertions, require cleanup, and increase the memory footprint.

\begin{figure}[tb]
    \centering
    \includegraphics[width=\columnwidth]{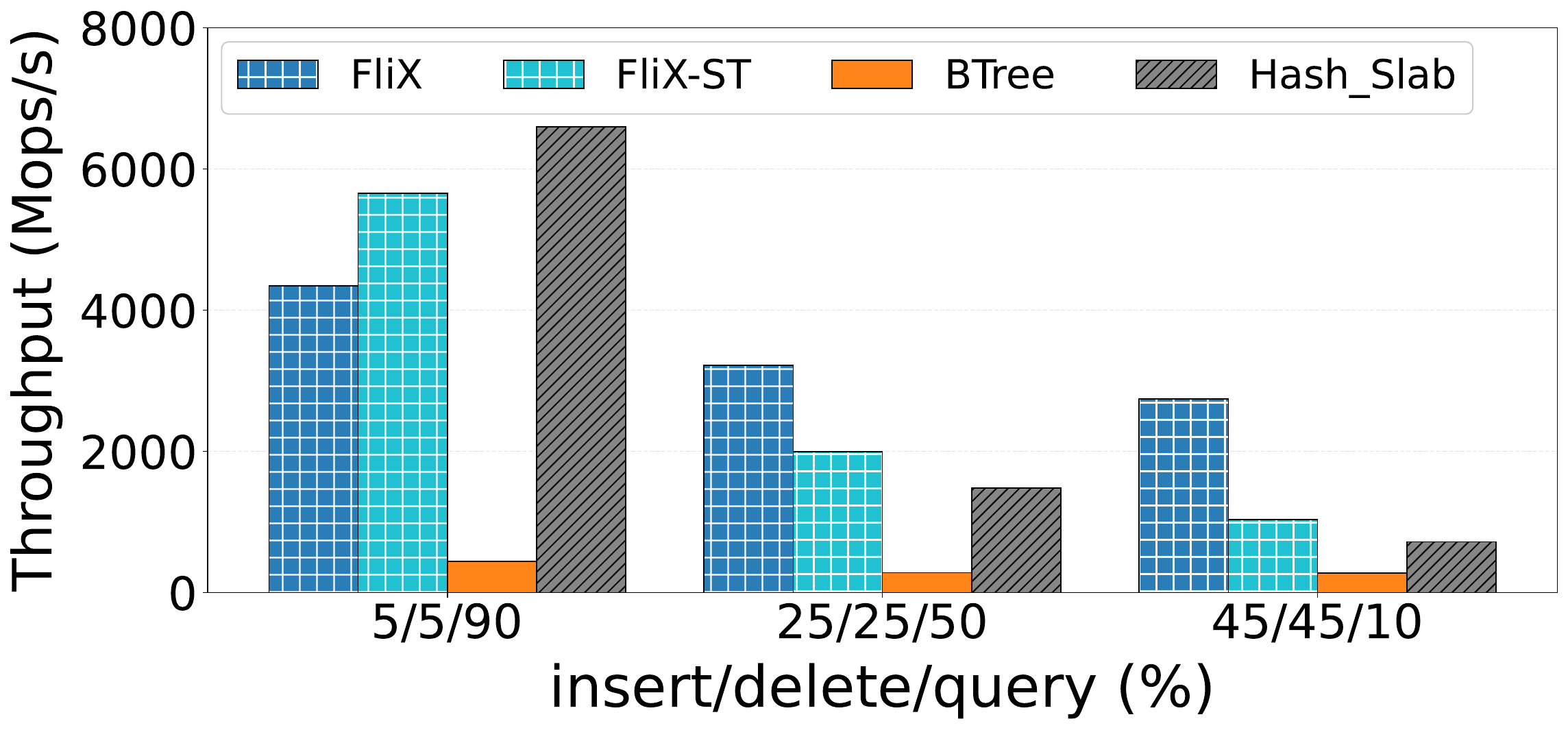}
    \caption{\small Fully concurrent kernels: \flix{} and \flixst{} against~\btree{} and~\slabhash{} on mixed batches of 100 million operations.}
    \label{fig:concurrent}
    \Description{Fully concurrent kernels}
\end{figure}

\subsection{Fully Concurrent Execution}
\label{sec:concurrent-updates-queries}

Next, we evaluate fully concurrent execution on mixed batches in which insertions, deletions, and queries are applied in a single kernel launch, using batches of 100 million operations per round. We compare~\flix{} with the~\btree{} and~\slabhash{}, the two baselines that support fully concurrent mixed batches. In Figure~\ref{fig:concurrent}, both~\flix{} implementations reclaim empty nodes on deletions and reuse them on insertions (shown with a grid pattern), whereas the~\btree{} and~\slabhash{} do not reclaim empty nodes during fully concurrent execution.

Figure~\ref{fig:concurrent} reports the average throughput across eight rounds for three operation mixes, $(5\%, 5\%, 90\%)$, $(25\%, 25\%, 50\%)$, and $(45\%, 45\%, 10\%)$, given as the percentages of insertions, deletions, and queries. \flix{} and~\flixst{} outperform the~\btree{} at every mix and overtake~\slabhash{} as the proportion of updates increases. For the update-heavy $(45\%, 45\%, 10\%)$ mix,~\flix{} achieves $4.42\times$ the throughput of~\slabhash{} and $11.39\times$ that of the~\btree{}.

\subsection{Restructuring~\flix{}}
\label{sec:restructuring-experiments}
We evaluate the effect of restructuring by allowing \flix{} to grow up to 300\% in 8 rounds of insertions followed by 8 rounds of deletions.
As shown in Table~\ref{tab:restructuring}, this process recovers up to 46\% of node space relative to the size of the data structure under a uniform distribution and has a positive impact on query performance.
The recovered space is obtained by merging underfull nodes.
This state is consistent with the original build configuration of \flix{}, allowing future insertions to proceed without the immediate need for further node splitting.

The impact of restructuring on query performance can be substantial. As indicated in Table~\ref{tab:restructuring}, we see up to a $4.35\times$ improvement in~\flixst{} query time when a restructuring operation is applied in round~8 following insertions. Query time remains at this lower threshold for all subsequent rounds. This translates to the following: The query time without restructuring averages $24$ ms from rounds~8 to~16 (X25Y90, 33M). Performing a restructuring operation in round~8, reduces the average query response time to $5.5$ ms. This is a significant change; however, we observe it in the~\flixst{} case. Further experimentation to determine ideal restructuring parameters is required for tiled queries in~\flix{}, since tiles benefit from more work assigned per bucket.  Restructuring immediately reduces the workload per bucket because the nodes in the data structure are flattened into one node/bucket.
Restructuring realigns the data structure after many updates have
occurred and costs approximately $20$--$50\times$ an update operation, depending on the rebuild frequency and build size.

\begin{table}[t]
\centering
\small
\begin{tabular}{L{1.5cm}@{\hspace{2mm}} l l r r r}
\hline
\textbf{Workload} & \textbf{Build} & \textbf{Final} & \textbf{Nodes } & \textbf{\%} & \textbf{Query} \\
\textbf{} & \textbf{size} & \textbf{size} & \textbf{recovered} & \textbf{recovery} & \textbf{speedup} \\
\hline
X25Y90 & 1M   & 4M   & 54,623    & 29\% & 1.01$\times$ \\
X25Y90 & 33M  & 134M & 1,745,428 & 29\% & 4.35$\times$ \\
X25Y90 & 100M & 300M & 3,928,746 & 28\% & 3.30$\times$ \\
X90Y90 & 1M   & 4M   & 81,307    & 43\% & 1.03$\times$ \\
X90Y90 & 33M  & 134M & 2,603,137 & 43\% & 3.65$\times$ \\
X90Y90 & 100M & 300M & 6,130,104 & 46\% & 1.57$\times$ \\
\hline
\end{tabular}

\caption{\small Node recovery through restructuring after successive phases of updates. Query speedup is the ratio of query response time before a restructuring operation to the time after it.}
\vspace{-2mm}
\label{tab:restructuring}
\Description{Rebuilding saving memory}

\end{table}

\subsection{Batched YCSB Workload}
\label{sec:ycsb}
\begin{figure}[tb!]
    \centering
    \setlength{\tabcolsep}{2pt}

    \begin{subfigure}[b]{0.49\linewidth}
        \centering
        \includegraphics[width=\linewidth]
        {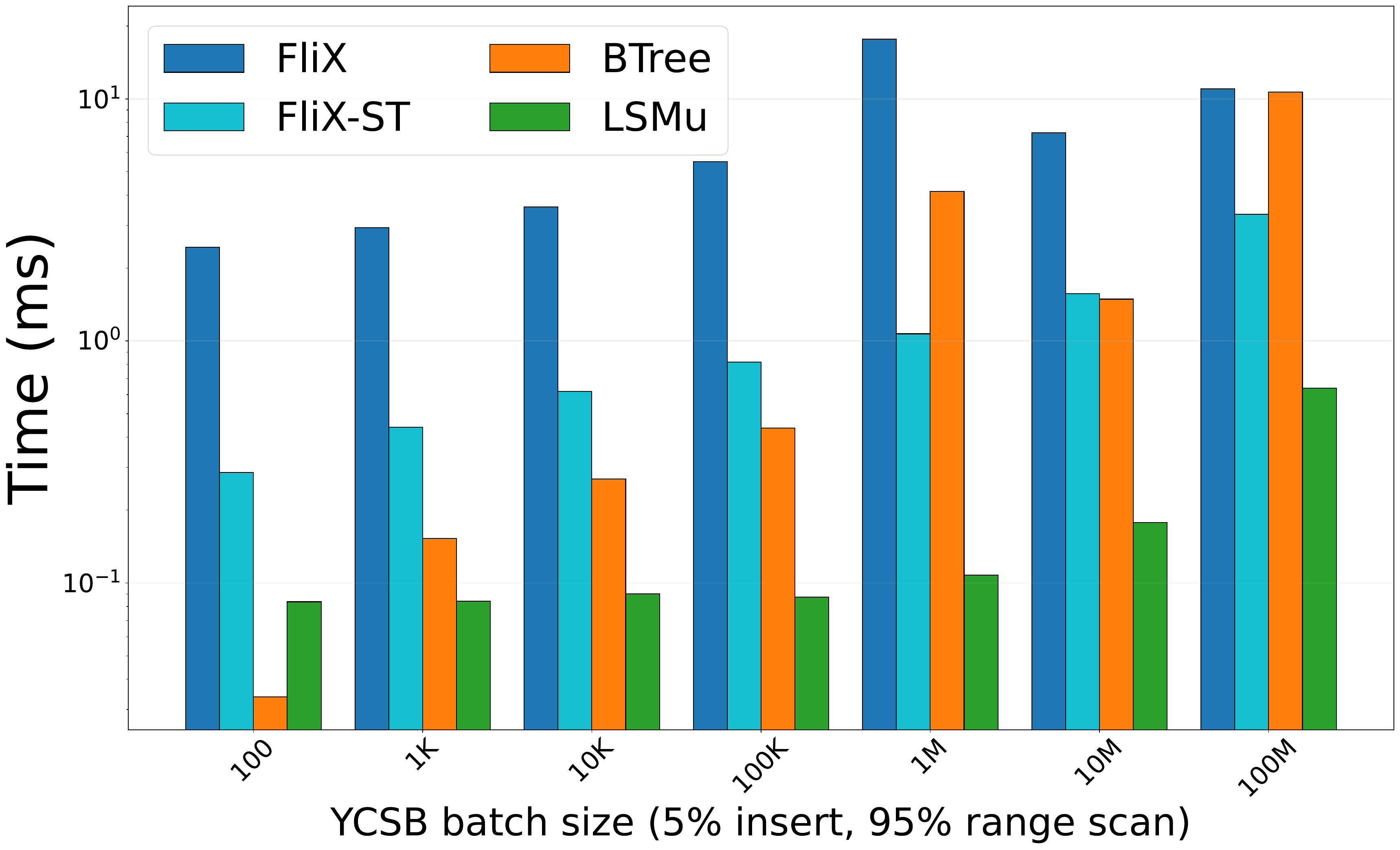}%
        \caption{Workload E (Inserts)}
    \end{subfigure}
    \hfill
    \begin{subfigure}[b]{0.49\linewidth}
        \centering
        \includegraphics[width=\linewidth]
        {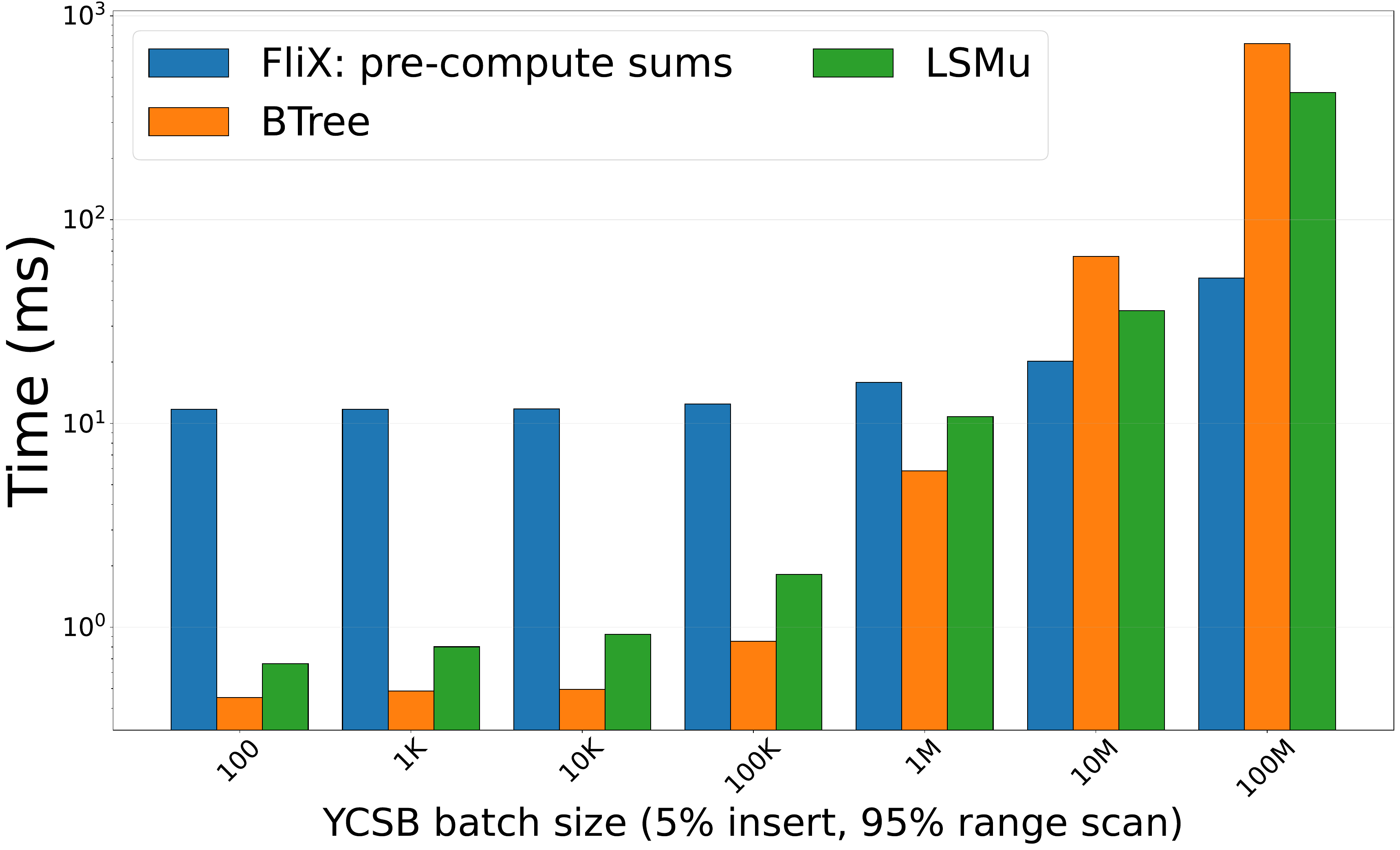}%
        \caption{Workload E (Range Scans)}
    \end{subfigure}

    \caption{\small Performance of batched GPU adaptations of
    YCSB Workload E (5\% insert \& 95\% range scans).
    }
    \vspace{-2mm}
    \label{fig:ycsb-workloads}
    \Description{Performance of batched GPU adaptations of YCSB
    Workloads E}

\end{figure}

Finally, we evaluate a batched adaptation of the Yahoo! Cloud Serving Benchmark (YCSB) for execution on the indexing structures tested. Operations are accumulated on the host and transferred to the GPU in batches.
We test across both ST and TL implementations of insertions in a YCSB-inspired workload~E.
Our experiments preserve the operation distributions and proportions defined by YCSB while adapting their execution to the batched GPU setting, and range aggregations. Workload-E performs (5\% $\times$ batch size) insertions and (95\% $\times$ batch size) range scans of lengths 1--100. The initial build size is 100 million uniform keys, range scans are drawn from a scattered Zipfian distribution across all live keys. For the~\flix{} implementation of range scans, we test the best-performing version, pre-computed sums.
Figure~\ref{fig:ycsb-workloads} (a) indicates that~\flixst{} insertion performance is stronger than the~\flix{} tiled insertions when the insertion workload is a smaller fraction of the batch size (5\%). Consistent with our results in Section~\ref{sec:update-experiments}, we find~\lsmu{} insertion performance is superior overall.~\flix{} pre-computed sums range scans outperform baselines when the range-scan batch size exceeds 10 million.

\section{Summary of Findings}
\label{sec:summary-of-findings}
Across a broad range of update workloads, the choice of update kernel must be workload-aware; tiled bulk kernels provide the best insertion and deletion performance under substantial growth ($\geq 200\%$) or highly skewed distributions, while single-threaded shift-right insertions remain highly competitive under lighter insertion pressure ($\leq 100\%$ growth).
\flix{} is competitive with \textit{unordered baselines} in query, insertion, and deletion performance; a workload consisting exclusively of point queries is still best served by a hash table, but the ordering maintained by~\flix{} comes without sacrificing point and range query performance, and even with the cost of sorting included,~\flix{} outperforms all baselines on large unsorted query batches. For range queries,~\flix{} is superior to its ordered competitors: \rqspan{} outperforms the~\btree{} and~\lsmu{} for almost all range sizes, and with precomputed aggregates,~\flix{} clearly outperforms both.
Although~\lsmu{} achieves strong update performance, its merge operations incur high memory overhead, and its deletions leave behind unconsolidated space. In contrast,~\flix{} makes more efficient use of GPU memory, surpasses~\lsmu{} in query performance, and is vastly superior in successor and range query performance.
Restructuring allows~\flix{} to reclaim a significant proportion of memory while improving query performance, and in fully concurrent execution,~\flix{} outperforms the~\btree{} and~\slabhash{} as the proportion of updates increases.

\vspace{2mm}

\section{Conclusion}
\label{sec:conclusions}

In this work, we presented~\flix{}, a fully GPU-resident indexing CDS that supports fast dynamic updates without sacrificing query performance.
By eliminating the traditional index layer and adopting a flipped indexing paradigm that maps compute to buckets,~\flix{} avoids many of the limitations of prior GPU indexing structures.
Our results show that \flix{} achieves the best overall performance among ordered competitors, while, as a secondary benefit, remaining surprisingly competitive with, and often superior to, unordered baselines such as hash tables.
In addition, \flix{} supports sustained long-running execution through GPU-resident restructuring and memory reclamation, and remains robust under substantial distributional shift in workloads.
Overall, these results suggest that flipped indexing could offer a practical, efficient, and scalable future for dynamic GPU database indexing.

\bibliographystyle{IEEEtran}
\bibliography{fliX}

@article{lit:rtindex,
  author       = {Justus Henneberg and
                  Felix Schuhknecht},
  title        = {{RTIndeX}: Exploiting Hardware-Accelerated {GPU} Raytracing for Database
                  Indexing},
  journal      = {Proc. {VLDB} Endow.},
  volume       = {16},
  number       = {13},
  pages        = {4268--4281},
  year         = {2023},
  url          = {https://www.vldb.org/pvldb/vol16/p4268-schuhknecht.pdf},
  bibsource    = {dblp computer science bibliography, https://dblp.org}
}

@inproceedings{lit:cgrxu,
  author={Henneberg, Justus and Schuhknecht, Felix and Kharal, Rosina and Brown, Trevor},
  booktitle={2025 IEEE 41st International Conference on Data Engineering (ICDE)}, 
  title={More Bang for Your Buck(et): Fast and Space-Efficient Hardware-Accelerated Coarse-Granular Indexing on GPUs}, 
  year={2025},
  volume={},
  number={},
  publisher = {IEEE},
  pages={1320-1333},
  address   = {Hong Kong, China},
  doi={10.1109/ICDE65448.2025.00103}
  }

@inproceedings{lit:gpuskiplist,
  title={A gpu-friendly skiplist algorithm},
  author={Moscovici, Nurit and Cohen, Nachshon and Petrank, Erez},
  booktitle={2017 26th International Conference on Parallel Architectures and Compilation Techniques (PACT)},
  pages={246--259},
  year={2017},
  number ={},
  address   = {Portland, OR, USA},
  publisher    = {IEEE},
  organization={IEEE}
}

@inproceedings{lit:warpcore,
  author       = {Daniel J{\"{u}}nger and
                  Robin Kobus and
                  Andr{\'{e}} M{\"{u}}ller and
                  Christian Hundt and
                  Kai Xu and
                  Weiguo Liu and
                  Bertil Schmidt},
  title        = {WarpCore: {A} Library for fast Hash Tables on GPUs},
  booktitle    = {27th {IEEE} International Conference on High Performance Computing,
                  Data, and Analytics, HiPC 2020, Pune, India, December 16-19, 2020},
  pages        = {11--20},
  publisher    = {{IEEE}},
  year         = {2020},
  address   = {Pune, India},
  url          = {https://doi.org/10.1109/HiPC50609.2020.00015},
  doi          = {10.1109/HIPC50609.2020.00015},
  bibsource    = {dblp computer science bibliography, https://dblp.org}
}

@inproceedings{lit:lsmtree,
  author    = {Saman Ashkiani and
               Shengren Li and
               Martin Farach{-}Colton and
               Nina Amenta and
               John D. Owens},
  title     = {{GPU} {LSM:} {A} Dynamic Dictionary Data Structure for the {GPU}},
  booktitle = {2018 {IEEE} International Parallel and Distributed Processing Symposium,
               {IPDPS} 2018, Vancouver, BC, Canada, May 21-25, 2018},
  pages     = {430--440},
  publisher = {{IEEE} Computer Society},
  address   = {N/A},
  year      = {2018},
  url       = {https://doi.org/10.1109/IPDPS.2018.00053},
  doi       = {10.1109/IPDPS.2018.00053},
  bibsource = {dblp computer science bibliography, https://dblp.org}
}

@inproceedings{lit:learnedindex,
  author    = {Xun Zhong and
               Yong Zhang and
               Yu Chen and
               Chao Li and
               Chunxiao Xing},
  title     = {Learned Index on {GPU}},
  booktitle = {38th {IEEE} International Conference on Data Engineering Workshops,
               {ICDE} Workshops 2022, Kuala Lumpur, Malaysia, May 9, 2022},
  pages     = {117--122},
  publisher = {{IEEE}},
  year      = {2022},
  url       = {https://doi.org/10.1109/ICDEW55742.2022.00024},
  doi       = {10.1109/ICDEW55742.2022.00024},
  bibsource = {dblp computer science bibliography, https://dblp.org}
}

@article{lit:hash-megakv,
  author    = {Kai Zhang and Kaibo Wang and Yuan Yuan and Lei Guo and Rubao Lee and Xiaodong Zhang},
  title     = {Mega-KV: {A} Case for GPUs to Maximize the Throughput of In-Memory Key-Value Stores},
  journal   = {Proc. {VLDB} Endow.},
  volume    = {8},
  number    = {11},
  pages     = {1226--1237},
  year      = {2015},
  url       = {http://www.vldb.org/pvldb/vol8/p1226-zhang.pdf},
  doi       = {10.14778/2809974.2809984}
}

@inproceedings{lit:btree1,
  author    = {Muhammad A. Awad and
               Saman Ashkiani and
               Rob Johnson and
               Martin Farach{-}Colton and
               John D. Owens},
  editor    = {Jeffrey K. Hollingsworth and
               Idit Keidar},
  title     = {Engineering a high-performance {GPU} B-Tree},
  booktitle = {Proceedings of the 24th {ACM} {SIGPLAN} Symposium on Principles and
               Practice of Parallel Programming, PPoPP 2019, Washington, DC, USA,
               February 16-20, 2019},
  pages     = {145--157},
  publisher = {{ACM}},
  address   = {N/A},
  year      = {2019},
  url       = {https://doi.org/10.1145/3293883.3295706},
  doi       = {10.1145/3293883.3295706},
  bibsource = {dblp computer science bibliography, https://dblp.org}
}

@inproceedings{lit:hash-slabhash,
  author    = {Saman Ashkiani and
               Martin Farach{-}Colton and
               John D. Owens},
  title     = {A Dynamic Hash Table for the {GPU}},
  booktitle = {2018 {IEEE} International Parallel and Distributed Processing Symposium,
               {IPDPS} 2018, Vancouver, BC, Canada, May 21-25, 2018},
  pages     = {419--429},
  publisher = {{IEEE} Computer Society},
  year      = {2018},
  address   = {N/A},
  url       = {https://doi.org/10.1109/IPDPS.2018.00052},
  doi       = {10.1109/IPDPS.2018.00052},
  bibsource = {dblp computer science bibliography, https://dblp.org}
}

@article{lit:hash-cudpp1,
  author    = {Dan A. Alcantara and
               Andrei Sharf and
               Fatemeh Abbasinejad and
               Shubhabrata Sengupta and
               Michael Mitzenmacher and
               John D. Owens and
               Nina Amenta},
  title     = {Real-time parallel hashing on the {GPU}},
  journal   = {{ACM} Trans. Graph.},
  volume    = {28},
  number    = {5},
  pages     = {154},
  year      = {2009},
  url       = {https://doi.org/10.1145/1618452.1618500},
  doi       = {10.1145/1618452.1618500},
  bibsource = {dblp computer science bibliography, https://dblp.org}
}

@article{lit:arch1,
  author    = {Periklis Chrysogelos and
               Manos Karpathiotakis and
               Raja Appuswamy and
               Anastasia Ailamaki},
  title     = {HetExchange: Encapsulating heterogeneous {CPU-GPU} parallelism in
               {JIT} compiled engines},
  journal   = {Proc. {VLDB} Endow.},
  volume    = {12},
  number    = {5},
  pages     = {544--556},
  year      = {2019},
  url       = {http://www.vldb.org/pvldb/vol12/p544-chrysogelos.pdf},
  doi       = {10.14778/3303753.3303760},
  bibsource = {dblp computer science bibliography, https://dblp.org}
}

@phdthesis{lit:awaddissertation,
  title={Fully Concurrent GPU Data Structures},
  author={Awad, Muhammad Abdelghaffar},
  year={2022},
  school={University of California, Davis}
}

@article{lit:fastinmemory,
  title={Fast In-Memory SQL Analytics on Typed Graphs.},
  author={Lin, Chunbin and Mandel, Benjamin and Papakonstantinou, Yannis and Springer, Matthias},
  journal={Proc. VLDB Endow.},
  volume={10},
  number={3},
  pages={265--276},
  year={2016}
}

@inproceedings{lit:mitgpucpu,
  title={A study of the fundamental performance characteristics of GPUs and CPUs for database analytics},
  author={Shanbhag, Anil and Madden, Samuel and Yu, Xiangyao},
  booktitle={Proceedings of the 2020 ACM SIGMOD international conference on Management of data},
  pages={1617--1632},
  year={2020},
  publisher = {Association for Computing Machinery},
  address   = {New York, NY, USA},
  doi       = {10.1145/3318464.3380595}
}

@inproceedings{lit:traditionalLSM,
  title     = {An Efficient Design and Implementation of LSM-tree Based Key-Value Store on Open-Channel SSD},
  author    = {Wang, Peng and Sun, Guangyu and Jiang, Song and Ouyang, Jian and Lin, Shiding and Zhang, Chen and Cong, Jason},
  booktitle = {Proceedings of the Ninth European Conference on Computer Systems},
  year      = {2014},
  publisher = {Association for Computing Machinery},
  address   = {New York, NY, USA},
  articleno = {16},
  numpages  = {14},
  doi       = {10.1145/2592798.2592804}
}

@inproceedings{lit:shahvarani2016hybrid,
  title={A hybrid B+-tree as solution for in-memory indexing on CPU-GPU heterogeneous computing platforms},
  author={Shahvarani, Amirhesam and Jacobsen, Hans-Arno},
 booktitle = {Proceedings of the 2016 International Conference on Management of Data},
  series    = {SIGMOD '16},
  pages     = {1523--1538},
  year = {2016},
  publisher={Association for Computing Machinery},
  address={New York, NY, USA}
}

@article{lit:hashfightlessley2020,
  title={HashFight: A platform-portable hash table for multi-core and many-core architectures},
  author={Lessley, Brenton and Li, Shaomeng and Childs, Hank},
  journal={Electronic Imaging},
  volume={32},
  pages={1--13},
  year={2020},
  publisher={Society for Imaging Science and Technology}
}

@unpublished{lit:coopgroupsharris2017,
  author  = {Harris, Mark and Perelygin, Kyrylo},
  title   = {Cooperative Groups: Flexible CUDA Thread Programming},
  date    = {2017-10-04},
  url     = {https://developer.nvidia.com/blog/cooperative-groups/},
  urldate = {2026-03-31},
  year = {2017},
  volume = {{\vphantom{1}}},
  organization = {NVIDIA Technical Blog}
}

@article{lit:boss,
  author       = {Hubert Mohr{-}Daurat and
                  Xuan Sun and
                  Holger Pirk},
  title        = {{BOSS} - An Architecture for Database Kernel Composition},
  journal      = {Proc. {VLDB} Endow.},
  volume       = {17},
  number       = {4},
  pages        = {877--890},
  year         = {2023},
  url          = {https://www.vldb.org/pvldb/vol17/p877-mohr-daurat.pdf},
  doi          = {10.14778/3636218.3636239},
  bibsource    = {dblp computer science bibliography, https://dblp.org}
}

@article{lit:maximus,
  author       = {Marko Kabic and
                  Shriram Chandran and
                  Gustavo Alonso},
  title        = {Maximus: {A} Modular Accelerated Query Engine for Data Analytics on
                  Heterogeneous Systems},
  journal      = {Proc. {ACM} Manag. Data},
  volume       = {3},
  number       = {3},
  pages        = {187:1--187:25},
  year         = {2025},
  url          = {https://doi.org/10.1145/3725324},
  doi          = {10.1145/3725324},
  bibsource    = {dblp computer science bibliography, https://dblp.org}
}

@inproceedings{lit:sirius,
  author       = {Bobbi Yogatama and
                  Yifei Yang and
                  Kevin Kristensen and
                  Devesh Sarda and
                  Abigale Kim and
                  Adrian Cockcroft and
                  Yu Teng and
                  Joshua Patterson and
                  Gregory Kimball and
                  Wes McKinney and
                  Weiwei Gong and
                  Xiangyao Yu},
  title        = {Rethinking Analytical Processing in the {GPU} Era},
  booktitle    = {16th Conference on Innovative Data Systems Research, {CIDR} 2026,
                  Chaminade, CA, USA, January 18-21, 2026},
  publisher    = {www.cidrdb.org},
  year         = {2026},
  url          = {https://vldb.org/cidrdb/2026/rethinking-analytical-processing-in-the-gpu-era.html},
  bibsource    = {dblp computer science bibliography, https://dblp.org}
}

@article{lit:gpuinterconnects,
  author       = {Marko Kabic and
                  Bowen Wu and
                  Jonas Dann and
                  Gustavo Alonso},
  title        = {Powerful GPUs or Fast Interconnects: Analyzing Relational Workloads
                  on Modern GPUs},
  journal      = {Proc. {VLDB} Endow.},
  volume       = {18},
  number       = {11},
  pages        = {4350--4363},
  year         = {2025},
  url          = {https://www.vldb.org/pvldb/vol18/p4350-kabic.pdf},
  doi          = {10.14778/3749646.3749698},
  bibsource    = {dblp computer science bibliography, https://dblp.org}
}

@inproceedings{lit:caldera,
  author       = {Raja Appuswamy and
                  Manos Karpathiotakis and
                  Danica Porobic and
                  Anastasia Ailamaki},
  title        = {The Case For Heterogeneous {HTAP}},
  booktitle    = {8th Biennial Conference on Innovative Data Systems Research, {CIDR}
                  2017, Chaminade, CA, USA, January 8-11, 2017, Online Proceedings},
  publisher    = {www.cidrdb.org},
  year         = {2017},
  url          = {http://cidrdb.org/cidr2017/papers/p21-appuswamy-cidr17.pdf},
  bibsource    = {dblp computer science bibliography, https://dblp.org}
}

@inproceedings{lit:gputestoftime,
  author       = {Aunn Raza and
                  Periklis Chrysogelos and
                  Panagiotis Sioulas and
                  Vladimir Indjic and
                  Angelos{-}Christos G. Anadiotis and
                  Anastasia Ailamaki},
  title        = {GPU-accelerated data management under the test of time},
  booktitle    = {10th Conference on Innovative Data Systems Research, {CIDR} 2020,
                  Amsterdam, The Netherlands, January 12-15, 2020, Online Proceedings},
  publisher    = {www.cidrdb.org},
  year         = {2020},
  url          = {https://vldb.org/cidrdb/2020/gpu-accelerated-data-management-under-the-test-of-time.html},
  bibsource    = {dblp computer science bibliography, https://dblp.org}
}

@inproceedings{ho1997range,
  author    = {Ching-Tien Ho and Rakesh Agrawal and Nimrod Megiddo and Ramakrishnan Srikant},
  title     = {Range Queries in {OLAP} Data Cubes},
  booktitle = {Proceedings of the 1997 ACM SIGMOD International Conference on Management of Data},
  pages     = {73--88},
  publisher = {ACM},
  year      = {1997},
  doi       = {10.1145/253260.253274}
}

\end{document}